\documentclass[10pt,journal,compsoc]{IEEEtran}
\usepackage{array}
\usepackage{makecell}
\usepackage{textcomp}
\usepackage{stfloats}
\usepackage{url}
\usepackage{verbatim}
\usepackage{graphicx}
\usepackage[caption=false,font=footnotesize,labelfont=rm,textfont=rm]{subfig}
\usepackage{float}
\usepackage{amsmath,amssymb,amsfonts}
\usepackage{mathrsfs}
\usepackage{epstopdf}
\usepackage[linesnumbered,algoruled,boxed,lined]{algorithm2e}
\usepackage[noend]{algpseudocode}
\usepackage[section]{placeins}
\allowdisplaybreaks[4]
\usepackage{multirow}
\usepackage{booktabs}
\usepackage{tabularx}
\usepackage{cleveref}
\usepackage{color}
\usepackage{bm}

\graphicspath{ {images/} }
\def\BibTeX{{\rm B\kern-.05em{\sc i\kern-.025em b}\kern-.08em
    T\kern-.1667em\lower.7ex\hbox{E}\kern-.125emX}}
\begin{document}

\title{Multi-AAV-enabled Distributed Beamforming in Low-Altitude Wireless Networking for AoI-Sensitive IoT Data Forwarding}

\author{Zifan Lang, Guixia Liu, Jiahui Li, 
        Geng Sun,~\IEEEmembership{Senior Member,~IEEE,}
        Zemin Sun, Jiacheng Wang,\\ 
        Dusit Niyato,~\IEEEmembership{Fellow,~IEEE,}
        Victor C.M. Leung,~\IEEEmembership{Life~Fellow,~IEEE}

\thanks{Zifan Lang, Guixia Liu, Jiahui Li, and Zemin Sun are with the College of Computer Science and Technology, Jilin University, Changchun 130012, China (e-mails: langzf23@mails.jlu.edu.cn; liugx@jlu.edu.cn; lijiahui@jlu.edu.cn; sunzemin@jlu.edu.cn).}
\thanks{Geng Sun is with the College of Computer Science and Technology, Jilin University, Changchun 130012, China, and with Key Laboratory of Symbolic Computation and Knowledge Engineering of Ministry of Education, Jilin University, Changchun 130012, China; he is also affiliated with the College of Computing and Data Science, Nanyang Technological University, Singapore 639798 (e-mail: sungeng@jlu.edu.cn). }
\thanks{Jiacheng Wang and Dusit Niyato are with the College of Computing and Data Science, Nanyang Technological University, Singapore (e-mails: jiacheng.wang@ntu.edu.sg; dniyato@ntu.edu.sg).}
\thanks{Victor C. M. Leung is with the Artificial Intelligence Research Institute, Shenzhen MSU-BIT University, Shenzhen 518115, China, also with the College of Computer Science and Software Engineering, Shenzhen University, Shenzhen 518060, China, and also with the Department of Electrical and Computer Engineering, The University of British Columbia, Vancouver V6T 1Z4, Canada (e-mail: vleung@ieee.org).}
\thanks{\textit{(Corresponding authors: Guixia Liu and Geng Sun.)}}
\thanks{Part of this paper appeared in IEEE ICC 2025~\cite{Lang2025}.}
}

\IEEEtitleabstractindextext{
    \begin{abstract}
    \par With the rapid development of low-altitude wireless networking (LAWN), autonomous aerial vehicles (AAVs) have emerged as critical enablers for timely and reliable data delivery, particularly in remote or underserved areas. In this context, the age of information (AoI) has emerged as a critical performance metric for evaluating the freshness and timeliness of transmitted information in Internet of things (IoT) networks. However, conventional AAV-assisted data transmission is fundamentally limited by finite communication coverage ranges, which requires periodic return flights for data relay operations. This propulsion-repositioning cycle inevitably introduces latency spikes that raise the AoI while degrading service reliability. To address these challenges, this paper proposes a AAV-assisted forwarding system based on distributed beamforming (DB) to enhance the AoI in IoT. Specifically, AAVs collaborate via distributed beamforming to collect and relay data between the sensor nodes (SNs) and remote base station (BS). Then, we formulate an optimization problem to minimize the AoI and AAV energy consumption, by jointly optimizing the AAV trajectories and communication schedules. Due to the non-convex nature of the problem and its pronounced temporal variability, we introduce a deep reinforcement learning solution that incorporates temporal sequence input, layer normalization gated recurrent unit, and a squeeze-and-excitation block to capture long-term dependencies, thereby improving decision-making stability and accuracy, and reducing computational complexity. Simulation results demonstrate that the proposed SAC-TLS algorithm outperforms baseline algorithms in terms of convergence, time average AoI, and energy consumption of AAVs. Moreover, extended simulation demonstrates that SAC-TLS maintains near-original performance with only a 20\% reduction in reward, even under 90\% model pruning, which confirms its suitability for resource-constrained AAV deployments.
    \end{abstract}
    
}
\maketitle

\IEEEpeerreviewmaketitle

\section{Introduction}
\IEEEPARstart{T}{he} rapid development of the low-altitude wireless networking (LAWN) has led to significant advancements in various sectors, particularly in communication and data transmission~\cite{Huang2024},~\cite{feng2025networked}. As autonomous aerial vehicles (AAVs) become integral to LAWN applications, they are increasingly being utilized to address critical challenges in remote areas, where traditional communication infrastructure is limited or nonexistent~\cite{He2024},~\cite{Xiao2024}. AAV-assisted data forwarding and relay systems have emerged as a key solution to extend the communication range, reduce the latency, and mitigate the obstacles such as signal attenuation and line-of-sight (LoS) obstructions~\cite{Sun2024f},~\cite{Xie2025}. By efficiently collecting, disseminating, and relaying data, AAVs significantly enhance the performance of Internet of things (IoT) networks in underserved and remote regions~\cite{Sun2024c}. 

\par Despite the advantages of AAV-assisted IoT, a significant challenge arises when timely data delivery is required~\cite{Sun2024d}. Specifically, the flight of AAVs introduces delays in data reception, which leads to a high age of information (AoI) and reduces real-time system responsiveness~\cite{Sun2024e},~\cite{Bai2025}. This issue becomes especially critical in the applications such as forest fire monitoring and border surveillance, where the delayed information results in severe consequences~\cite{Sun2024b}. Therefore, minimizing the AAV flight time while maintaining efficient data transmission to reduce AoI remains an urgent task. Addressing this challenge demands innovative approaches that ensure both timely data delivery and efficient resource utilization.

\par Distributed beamforming (DB) offers a promising solution to these challenges by significantly enhancing the transmission gain and extending the communication range~\cite{Li2024}, \cite{Li2024c}, \cite{Zhang2024}. By employing distributed beamforming, AAVs can maintain connectivity with the BS without frequent flights, thereby mitigating flight-induced delays and reducing AoI~\cite{Li2023},~\cite{Liu2024}. This capability is particularly valuable in the scenarios requiring continuous data relay, as it allows for more efficient use of AAV energy resources.

\par Designing an effective and reliable AAV-assisted relay system involves several complexities. First, distributed beamforming introduces multiple transmission phases, which increases the difficulty of system modeling. Additionally, optimizing AAV trajectories is essential for minimizing both AoI and energy consumption across sensors. Furthermore, the uncertainty and variability introduced by dynamic updates of sensor data demand robust optimization strategies that can adapt in real time. However, addressing these complexities requires innovative modeling and optimization techniques that can effectively handle the dynamic and multi-faceted nature of the problem.

\par Unlike previous studies, this paper introduces a AAV-assisted AoI-sensitive data forwarding system and proposes a modified deep reinforcement learning (DRL) algorithm to minimize the AoI of SNs and energy consumption of AAVs. The primary contributions of this work are outlined as follows:

\begin{itemize}
    \item \textit{AAV-assisted AoI-sensitive Data Forwarding System:}
    We consider a AAV-assisted AoI-sensitive data forwarding system with distributed beamforming. Specifically, the direct link from the SNs to BS is blocked due to the long-range transmission distance, and distributed beamforming is employed to mitigate flight-induced delays of AAVs. To the best of our knowledge, such a data forwarding system with distributed beamforming to reduce AoI has not yet been investigated in the literature. 
    \item \textit{Formulation of Complex Optimization Problem:}
    In the considered system, we formulate a joint optimization problem that cooperatively plans the AAV trajectory and schedules communication to minimize the AoI of SNs and the energy consumption of AAVs. The real-time requirements and dynamic nature of this problem make it challenging to be solved efficiently.
    \item \textit{Enhanced DRL-based Approach:}
    We propose a soft actor-critic with temporal sequence, layer normalization gated recurrent unit, and squeeze-and-excitation block (SAC-TLS), which is a DRL-based algorithm to solve the formulated optimization problem. Specifically, temporal sequence improves the learning capability of standard SAC. Layer normalization gated recurrent unit (LNGRU) captures long-term dependencies in sequential data. Moreover, we introduce squeeze-and-excitation (SE) block to enhance the focus of SAC on critical elements with input sequences. 
    \item \textit{Simulation and Performance Evaluation:}
    Simulation results verify that SAC-TLS outperforms all baseline algorithms by reducing the average AoI by 17.3\% and increasing cumulative rewards by 24.5\%. The algorithm converges 1.6 times faster due to effective temporal sequence modeling and LNGRU stabilization. Ablation studies further confirm the critical role of LNGRU and SE block in enhancing learning stability and decision accuracy. Furthermore, compression results demonstrate that SAC-TLS maintains near-original performance with 20\% reward reduction at 90\% pruning under optimal regularization. 
\end{itemize}

\par The remainder of this paper is structured as follows. Section~\ref{sec:related work} provides a review of key related works. Section~\ref{sec:system model} describes the system models and preliminaries. The optimization problem is formulated in Section~\ref{sec:problem formulation}. Moreover, Section~\ref{sec:SAC-TLS} details the proposed SAC-TLS algorithm. Simulation results are discussed in Section~\ref{sec:simulation}. Finally, conclusions are drawn in Section~\ref{sec:conclusion}.

\section{Related Work}\label{sec:related work}
\par In our work, we aim to propose a AAV-assisted AoI-sensitive data forwarding system with distributed beamforming by optimizing the AAV trajectory and communication schedules. In the following, we will review several key related works to highlight the novelty of our research. Furthermore, Table \ref{tab:comparison} summarizes the main distinctions between these works and this work. 

\begin{table*}[]
\centering
\caption{Comparison Between Related Works With This Work}
\label{tab:comparison}
\resizebox{\textwidth}{!}{
\begin{tabular}{cccccccccc}
\hline
\multicolumn{1}{|c|}{\multirow{3}{*}{\textbf{Reference}}}            & \multicolumn{2}{c|}{ \textbf{System model}}           & \multicolumn{2}{c|}{ \textbf{Optimization objective}}   & \multicolumn{2}{c|}{ \textbf{Optimization variable}}                      & \multicolumn{3}{c|}{ \textbf{Optimization methods}}                                                          \\ \cline{2-10} 
\multicolumn{1}{|c|}{}                                      & \multicolumn{1}{c|}{\multirow{2}{*}{\textbf{Multi-AAV}}}    & \multicolumn{1}{c|}{\multirow{2}{*}{\textbf{DB}}}  & \multicolumn{1}{c|}{\textbf{AAV}}          & \multicolumn{1}{c|}{\textbf{Communication}} & \multicolumn{1}{c|}{\multirow{2}{*}{\textbf{AoI}}} & \multicolumn{1}{c|}{\multirow{2}{*}{\textbf{Energy}}} & \multicolumn{1}{c|}{\multirow{2}{*}{\textbf{DRL}}} & \multicolumn{1}{c|}{\textbf{Temporal}} & \multicolumn{1}{c|}{\textbf{Feature}} \\
\multicolumn{1}{|c|}{}                                      & \multicolumn{1}{c|}{} & \multicolumn{1}{c|}{}  & \multicolumn{1}{c|}{\textbf{trajectory}}   & \multicolumn{1}{c|}{\textbf{scheduling}}    & \multicolumn{1}{c|}{}                     & \multicolumn{1}{c|}{}                        & \multicolumn{1}{c|}{}                     & \multicolumn{1}{c|}{\textbf{dependency}}                  & \multicolumn{1}{c|}{\textbf{recalibration}}                  \\ \hline
\multicolumn{1}{|c|}{\cite{qin2022aoi}}        & \multicolumn{1}{c|}{$\times$}     & \multicolumn{1}{c|}{$\times$}     & \multicolumn{1}{c|}{$\checkmark$} & \multicolumn{1}{c|}{$\checkmark$}  & \multicolumn{1}{c|}{$\checkmark$}         & \multicolumn{1}{c|}{$\times$}                & \multicolumn{1}{c|}{$\times$}             & \multicolumn{1}{c|}{$\checkmark$}                  & \multicolumn{1}{c|}{$\times$}                  \\ \hline
\multicolumn{1}{|c|}{\cite{Zhu2023}}           & \multicolumn{1}{c|}{$\times$}     & \multicolumn{1}{c|}{$\times$}     & \multicolumn{1}{c|}{$\checkmark$} & \multicolumn{1}{c|}{$\times$}      & \multicolumn{1}{c|}{$\checkmark$}         & \multicolumn{1}{c|}{$\times$}                & \multicolumn{1}{c|}{$\times$}             & \multicolumn{1}{c|}{$\checkmark$}                  & \multicolumn{1}{c|}{$\times$}                  \\ \hline
\multicolumn{1}{|c|}{\cite{huang2025learning}} & \multicolumn{1}{c|}{$\times$}     & \multicolumn{1}{c|}{$\times$}     & \multicolumn{1}{c|}{$\checkmark$} & \multicolumn{1}{c|}{$\times$}      & \multicolumn{1}{c|}{$\checkmark$}         & \multicolumn{1}{c|}{$\times$}                & \multicolumn{1}{c|}{$\checkmark$}         & \multicolumn{1}{c|}{$\checkmark$}                  & \multicolumn{1}{c|}{$\times$}                  \\ \hline
\multicolumn{1}{|c|}{\cite{yang2025oh}}        & \multicolumn{1}{c|}{$\times$}     & \multicolumn{1}{c|}{$\times$}     & \multicolumn{1}{c|}{$\checkmark$} & \multicolumn{1}{c|}{$\times$}      & \multicolumn{1}{c|}{$\checkmark$}         & \multicolumn{1}{c|}{$\times$}                & \multicolumn{1}{c|}{$\checkmark$}         & \multicolumn{1}{c|}{$\checkmark$}                  & \multicolumn{1}{c|}{$\times$}                  \\ \hline
\multicolumn{1}{|c|}{\cite{zhou2025}}        & \multicolumn{1}{c|}{$\times$}     & \multicolumn{1}{c|}{$\times$}     & \multicolumn{1}{c|}{$\checkmark$} & \multicolumn{1}{c|}{$\checkmark$}      & \multicolumn{1}{c|}{$\checkmark$}         & \multicolumn{1}{c|}{$\times$}                & \multicolumn{1}{c|}{$\times$}         & \multicolumn{1}{c|}{$\times$}                  & \multicolumn{1}{c|}{$\checkmark$}    \\ \hline
\multicolumn{1}{|c|}{\cite{han2022age}}        & \multicolumn{1}{c|}{$\checkmark$} & \multicolumn{1}{c|}{$\times$}     & \multicolumn{1}{c|}{$\times$}     & \multicolumn{1}{c|}{$\times$}      & \multicolumn{1}{c|}{$\checkmark$}         & \multicolumn{1}{c|}{$\times$}                & \multicolumn{1}{c|}{$\times$}             & \multicolumn{1}{c|}{$\checkmark$}                  & \multicolumn{1}{c|}{$\times$}                  \\ \hline
\multicolumn{1}{|c|}{\cite{Yang2023}}          & \multicolumn{1}{c|}{$\checkmark$} & \multicolumn{1}{c|}{$\times$}     & \multicolumn{1}{c|}{$\times$}     & \multicolumn{1}{c|}{$\times$}      & \multicolumn{1}{c|}{$\checkmark$}         & \multicolumn{1}{c|}{$\times$}                & \multicolumn{1}{c|}{$\times$}             & \multicolumn{1}{c|}{$\checkmark$}                  & \multicolumn{1}{c|}{$\times$}                  \\ \hline
\multicolumn{1}{|c|}{\cite{Feng2024}}          & \multicolumn{1}{c|}{$\checkmark$} & \multicolumn{1}{c|}{$\times$}     & \multicolumn{1}{c|}{$\times$}     & \multicolumn{1}{c|}{$\checkmark$}  & \multicolumn{1}{c|}{$\checkmark$}         & \multicolumn{1}{c|}{$\times$}                & \multicolumn{1}{c|}{$\times$}             & \multicolumn{1}{c|}{$\checkmark$}                  & \multicolumn{1}{c|}{$\times$}                  \\ \hline
\multicolumn{1}{|c|}{\cite{Zhan2024a}}          & \multicolumn{1}{c|}{$\checkmark$} & \multicolumn{1}{c|}{$\times$}     & \multicolumn{1}{c|}{$\checkmark$}     & \multicolumn{1}{c|}{$\times$}  & \multicolumn{1}{c|}{$\checkmark$}         & \multicolumn{1}{c|}{$\times$}                & \multicolumn{1}{c|}{$\times$}             & \multicolumn{1}{c|}{$\checkmark$}                  & \multicolumn{1}{c|}{$\times$}                  \\ \hline
\multicolumn{1}{|c|}{\cite{long2025lyapunov}}  & \multicolumn{1}{c|}{$\checkmark$} & \multicolumn{1}{c|}{$\times$}     & \multicolumn{1}{c|}{$\checkmark$} & \multicolumn{1}{c|}{$\times$}      & \multicolumn{1}{c|}{$\checkmark$}         & \multicolumn{1}{c|}{$\times$}                & \multicolumn{1}{c|}{$\checkmark$}         & \multicolumn{1}{c|}{$\times$}                  & \multicolumn{1}{c|}{$\checkmark$}                  \\ \hline
\multicolumn{1}{|c|}{\cite{Han2022a}}          & \multicolumn{1}{c|}{$\checkmark$} & \multicolumn{1}{c|}{$\times$}     & \multicolumn{1}{c|}{$\checkmark$} & \multicolumn{1}{c|}{$\times$}      & \multicolumn{1}{c|}{$\checkmark$}         & \multicolumn{1}{c|}{$\times$}                & \multicolumn{1}{c|}{$\times$}             & \multicolumn{1}{c|}{$\checkmark$}                  & \multicolumn{1}{c|}{$\times$}                  \\ \hline
\multicolumn{1}{|c|}{\cite{Dai2022}}          & \multicolumn{1}{c|}{$\checkmark$} & \multicolumn{1}{c|}{$\times$}     & \multicolumn{1}{c|}{$\checkmark$} & \multicolumn{1}{c|}{$\times$}      & \multicolumn{1}{c|}{$\checkmark$}         & \multicolumn{1}{c|}{$\times$}                & \multicolumn{1}{c|}{$\checkmark$}             & \multicolumn{1}{c|}{$\times$}                  & \multicolumn{1}{c|}{$\checkmark$}                  \\ \hline
\multicolumn{1}{|c|}{\cite{Zheng2024}}        & \multicolumn{1}{c|}{$\checkmark$} & \multicolumn{1}{c|}{$\checkmark$} & \multicolumn{1}{c|}{$\checkmark$} & \multicolumn{1}{c|}{$\times$}      & \multicolumn{1}{c|}{$\times$}             & \multicolumn{1}{c|}{$\checkmark$}            & \multicolumn{1}{c|}{$\times$}             & \multicolumn{1}{c|}{$\times$}                  & \multicolumn{1}{c|}{$\checkmark$}           \\ \hline
\multicolumn{1}{|c|}{\cite{liu2025joint}}        & \multicolumn{1}{c|}{$\times$}     & \multicolumn{1}{c|}{$\times$}     & \multicolumn{1}{c|}{$\checkmark$} & \multicolumn{1}{c|}{$\times$}      & \multicolumn{1}{c|}{$\checkmark$}         & \multicolumn{1}{c|}{$\times$}                & \multicolumn{1}{c|}{$\times$}         & \multicolumn{1}{c|}{$\times$}                  & \multicolumn{1}{c|}{$\times$}                  \\ \hline
\multicolumn{1}{|c|}{\cite{Liu2025}}        & \multicolumn{1}{c|}{$\times$} & \multicolumn{1}{c|}{$\times$} & \multicolumn{1}{c|}{$\checkmark$} & \multicolumn{1}{c|}{$\times$}      & \multicolumn{1}{c|}{$\checkmark$}             & \multicolumn{1}{c|}{$\times$}            & \multicolumn{1}{c|}{$\times$}             & \multicolumn{1}{c|}{$\checkmark$}                  & \multicolumn{1}{c|}{$\times$} \\ \hline
\multicolumn{1}{|c|}{\cite{Samir2022}}         & \multicolumn{1}{c|}{$\checkmark$} & \multicolumn{1}{c|}{$\times$}     & \multicolumn{1}{c|}{$\times$}     & \multicolumn{1}{c|}{$\checkmark$}  & \multicolumn{1}{c|}{$\checkmark$}         & \multicolumn{1}{c|}{$\times$}                & \multicolumn{1}{c|}{$\checkmark$}         & \multicolumn{1}{c|}{$\checkmark$}                  & \multicolumn{1}{c|}{$\times$}                  \\ \hline
\multicolumn{1}{|c|}{\cite{Zakeri2024}}         & \multicolumn{1}{c|}{$\times$} & \multicolumn{1}{c|}{$\times$}     & \multicolumn{1}{c|}{$\times$}     & \multicolumn{1}{c|}{$\checkmark$}  & \multicolumn{1}{c|}{$\checkmark$}         & \multicolumn{1}{c|}{$\times$}                & \multicolumn{1}{c|}{$\checkmark$}         & \multicolumn{1}{c|}{$\checkmark$}                  & \multicolumn{1}{c|}{$\times$}                  \\ \hline
\multicolumn{1}{|c|}{\cite{Qu2024}}         & \multicolumn{1}{c|}{$\times$} & \multicolumn{1}{c|}{$\times$}     & \multicolumn{1}{c|}{$\times$}     & \multicolumn{1}{c|}{$\checkmark$}  & \multicolumn{1}{c|}{$\checkmark$}         & \multicolumn{1}{c|}{$\times$}                & \multicolumn{1}{c|}{$\times$}         & \multicolumn{1}{c|}{$\times$}                  & \multicolumn{1}{c|}{$\times$}                  \\ \hline
\multicolumn{1}{|c|}{\cite{Zhao2025}}          & \multicolumn{1}{c|}{$\times$}     & \multicolumn{1}{c|}{$\times$}     & \multicolumn{1}{c|}{$\times$}     & \multicolumn{1}{c|}{$\checkmark$}  & \multicolumn{1}{c|}{$\checkmark$}         & \multicolumn{1}{c|}{$\times$}                & \multicolumn{1}{c|}{$\times$}             & \multicolumn{1}{c|}{$\times$}                  & \multicolumn{1}{c|}{$\checkmark$}    \\ \hline
\multicolumn{1}{|c|}{\cite{chen2025average}}          & \multicolumn{1}{c|}{$\times$}     & \multicolumn{1}{c|}{$\times$}     & \multicolumn{1}{c|}{$\times$}     & \multicolumn{1}{c|}{$\checkmark$}  & \multicolumn{1}{c|}{$\checkmark$}         & \multicolumn{1}{c|}{$\times$}                & \multicolumn{1}{c|}{$\times$}             & \multicolumn{1}{c|}{$\checkmark$}          & \multicolumn{1}{c|}{$\times$}    \\ \hline
\multicolumn{1}{|c|}{\cite{Gao2023}}           & \multicolumn{1}{c|}{$\checkmark$} & \multicolumn{1}{c|}{$\times$}     & \multicolumn{1}{c|}{$\checkmark$} & \multicolumn{1}{c|}{$\times$}      & \multicolumn{1}{c|}{$\checkmark$}         & \multicolumn{1}{c|}{$\times$}                & \multicolumn{1}{c|}{$\times$}             & \multicolumn{1}{c|}{$\checkmark$}                  & \multicolumn{1}{c|}{$\times$}                  \\ \hline
\multicolumn{1}{|c|}{\cite{Qi2023}}            & \multicolumn{1}{c|}{$\checkmark$} & \multicolumn{1}{c|}{$\times$}     & \multicolumn{1}{c|}{$\checkmark$} & \multicolumn{1}{c|}{$\times$}      & \multicolumn{1}{c|}{$\checkmark$}         & \multicolumn{1}{c|}{$\times$}                & \multicolumn{1}{c|}{$\times$}             & \multicolumn{1}{c|}{$\times$}                  & \multicolumn{1}{c|}{$\checkmark$}                  \\ \hline
\multicolumn{1}{|c|}{\cite{Hoang2024}}         & \multicolumn{1}{c|}{$\times$}     & \multicolumn{1}{c|}{$\times$}     & \multicolumn{1}{c|}{$\checkmark$} & \multicolumn{1}{c|}{$\times$}      & \multicolumn{1}{c|}{$\checkmark$}         & \multicolumn{1}{c|}{$\times$}                & \multicolumn{1}{c|}{$\times$}             & \multicolumn{1}{c|}{$\times$}                  & \multicolumn{1}{c|}{$\checkmark$}                  \\ \hline
\multicolumn{1}{|c|}{\cite{Zhan2024}}          & \multicolumn{1}{c|}{$\times$}     & \multicolumn{1}{c|}{$\times$}     & \multicolumn{1}{c|}{$\checkmark$} & \multicolumn{1}{c|}{$\checkmark$}  & \multicolumn{1}{c|}{$\checkmark$}         & \multicolumn{1}{c|}{$\times$}                & \multicolumn{1}{c|}{$\times$}             & \multicolumn{1}{c|}{$\checkmark$}                  & \multicolumn{1}{c|}{$\times$}                  \\ \hline
\multicolumn{1}{|c|}{\cite{Reddy2025}}        & \multicolumn{1}{c|}{$\times$}     & \multicolumn{1}{c|}{$\times$}     & \multicolumn{1}{c|}{$\times$}     & \multicolumn{1}{c|}{$\checkmark$}  & \multicolumn{1}{c|}{$\checkmark$}         & \multicolumn{1}{c|}{$\times$}            & \multicolumn{1}{c|}{$\times$}         & \multicolumn{1}{c|}{$\checkmark$}         & \multicolumn{1}{c|}{$\times$}    \\ \hline
\multicolumn{1}{|c|}{\cite{Zheng2024a}}        & \multicolumn{1}{c|}{$\times$}     & \multicolumn{1}{c|}{$\times$}     & \multicolumn{1}{c|}{$\times$}     & \multicolumn{1}{c|}{$\checkmark$}  & \multicolumn{1}{c|}{$\checkmark$}         & \multicolumn{1}{c|}{$\checkmark$}            & \multicolumn{1}{c|}{$\checkmark$}         & \multicolumn{1}{c|}{$\times$}                  & \multicolumn{1}{c|}{$\times$}                  \\ \hline
\multicolumn{1}{|c|}{\cite{Li2024b}}           & \multicolumn{1}{c|}{$\checkmark$} & \multicolumn{1}{c|}{$\times$}     & \multicolumn{1}{c|}{$\checkmark$} & \multicolumn{1}{c|}{$\times$}      & \multicolumn{1}{c|}{$\checkmark$}         & \multicolumn{1}{c|}{$\checkmark$}            & \multicolumn{1}{c|}{$\checkmark$}         & \multicolumn{1}{c|}{$\checkmark$}                  & \multicolumn{1}{c|}{$\times$}                  \\ \hline
\multicolumn{1}{|c|}{\cite{zhang2025aoi}}      & \multicolumn{1}{c|}{$\checkmark$} & \multicolumn{1}{c|}{$\times$}     & \multicolumn{1}{c|}{$\checkmark$} & \multicolumn{1}{c|}{$\times$}      & \multicolumn{1}{c|}{$\checkmark$}         & \multicolumn{1}{c|}{$\times$}                & \multicolumn{1}{c|}{$\checkmark$}         & \multicolumn{1}{c|}{$\times$}                  & \multicolumn{1}{c|}{$\times$}                  \\ \hline
\multicolumn{1}{|c|}{\cite{Huang2025}}         & \multicolumn{1}{c|}{$\times$}     & \multicolumn{1}{c|}{$\times$}     & \multicolumn{1}{c|}{$\checkmark$} & \multicolumn{1}{c|}{$\times$}      & \multicolumn{1}{c|}{$\checkmark$}         & \multicolumn{1}{c|}{$\checkmark$}            & \multicolumn{1}{c|}{$\checkmark$}         & \multicolumn{1}{c|}{$\times$}                  & \multicolumn{1}{c|}{$\times$}                  \\ \hline
\multicolumn{1}{|c|}{\cite{Zhang2025}}         & \multicolumn{1}{c|}{$\times$}     & \multicolumn{1}{c|}{$\times$}     & \multicolumn{1}{c|}{$\times$} & \multicolumn{1}{c|}{$\checkmark$}      & \multicolumn{1}{c|}{$\checkmark$}         & \multicolumn{1}{c|}{$\times$}            & \multicolumn{1}{c|}{$\checkmark$}         & \multicolumn{1}{c|}{$\times$}                  & \multicolumn{1}{c|}{$\checkmark$}                  \\ \hline
\multicolumn{1}{|c|}{This work}                                                   & \multicolumn{1}{c|}{$\checkmark$}                      & \multicolumn{1}{c|}{$\checkmark$}                       & \multicolumn{1}{c|}{$\checkmark$}                     & \multicolumn{1}{c|}{$\checkmark$}                      & \multicolumn{1}{c|}{$\checkmark$}                               & \multicolumn{1}{c|}{$\checkmark$}                                 & \multicolumn{1}{c|}{$\checkmark$}                               & \multicolumn{1}{c|}{$\checkmark$}                           & \multicolumn{1}{c|}{$\checkmark$}                           \\ \hline
\end{tabular}}
\end{table*}

\subsection{AoI in AAV-Assisted Communication}
\par AoI has become a critical metric in AAV-assisted communication systems, as it quantifies the freshness of information, which is particularly important for urgent response scenarios such as disaster response and real-time surveillance. 
Recent research has demonstrated that optimizing AoI can significantly enhance the performance of AAV-assisted IoT networks. For instance, Qin \textit{et al.}~\cite{qin2022aoi} investigated the AoI-aware scheduling for air-ground collaborative mobile edge computing, they optimized task offloading between the AAVs and ground servers to minimize system-wide AoI. Zhu \textit{et al.}\cite{Zhu2023} considered a cluster-based IoT network. They addressed the AAV trajectory planning and visit sequence scheduling to ensure the freshness of the collected information. As illustrated in~\cite{huang2025learning}, Huang \textit{et al.} investigated AoI-optimal trajectory planning in AAV-assisted IoT networks. They proposed a framework where the AAV switches between the flying and hovering modes to collect data from widely distributed IoT devices. Moreover, Yang \textit{et al.}~\cite{yang2025oh} introduced a AAV-assisted IoT data collection system, which incorporates an energy-efficient simultaneously transmitting and reflecting reconfigurable intelligent surface (STAR-RIS) beamforming framework. 
Zhou \textit{et al.}~\cite{zhou2025} addressed reliability challenges in AAV-assisted mobile edge computing systems within space-air-ground integrated networks by developing a stochastic modeling approach that accounts for variable communication channels, data loads, and node mobility. The authors proposed a comprehensive reliability optimization framework that integrates resource allocation, scheduling, and AAV trajectory. 
However, these studies above adopt single-AAV-assisted AoI optimization methods, and fail to consider multi-AAV collaboration or real-time adaptability in dynamic environments. 

\par Recently, the multi-AAV collaborative system demonstrates performance advantages that significantly surpass single AAV and conventional ground communication architecture through dynamic task allocation, distributed resource coordination and collaborative signal enhancement. Specifically, Han \textit{et al.}~\cite{han2022age} proposed a game-theoretic framework for AAV-aided vehicular edge computing networks, which reduced AoI and improved energy efficiency through cooperative resource allocation. They modeled multi-agent interactions between the AAVs and vehicles under static node assumptions. Yang \textit{et al.}\cite{Yang2023} investigated AAV-aided traffic monitoring networks under adversarial conditions to minimize the AoI in the presence of malicious attackers. Likewise, Feng \textit{et al.}~\cite{Feng2024} evaluated the performance of AAV-aided wireless networks using AoI as a key metric. They compared two communication scenarios which are AAV-to-AAV (U2U) and AAV-to-network (U2N), and analyzed the impact of hybrid automatic repeat request (HARQ) protocols on the timeliness of data transmission. Zhan \textit{et al.}~\cite{Zhan2024a} explored online design policies for cellular-connected multiple AAV communications without prior channel condition knowledge. They focused on wireless resource allocation and 3D path planning to maximize minimum uplink throughput while respecting AAV energy constraints. 
In~\cite{long2025lyapunov}, Long \textit{et al.} introduced a novel semantic-aware AoI (SAoI) metric, which balances the information freshness and semantic value, and formulated a time-averaged SAoI minimization problem. They decoupled the problem by using the Lyapunov framework and applied hierarchical DRL to optimize AAV-ground user associations, semantic extraction, and AAV trajectories. However, they ignore the complexities and challenges posed by remote and dynamic environments, where communication conditions can be highly unpredictable. 

\par However, the existing AAV-assisted communication systems to optimize AoI face limitations in maintaining efficient data transmission over large-scale and dynamic environments. To this end, we propose an enhanced AAV-assisted AoI optimization system with distributed beamforming to address the shortcomings of the existing research. 

\subsection{Optimizations for AAV Trajectory and Communication Scheduling}
\par Recent studies have demonstrated that optimizing the trajectory of AAV can significantly enhance the freshness of information and improve the energy efficiency of AAV-assisted systems. For instance, Han \textit{et al.}~\cite{Han2022a} explored the use of AAVs in intelligent transportation systems to enhance data collection and provide low-latency vehicular services. The study analyzed the performance of AAV-aided IoT by optimizing the deployment of multiple AAVs to minimize the average peak AoI, considering traffic intensity, seamless coverage, finite queue, and coverage probability constraints. 
In~\cite{Dai2022}, Dai \textit{et al.} addressed the challenge of optimizing AAV trajectory in mobile crowd sensing scenarios. They proposed a DRL-based framework to maximizie the volume of collected data and minimize the AoI while ensuring broad geographical coverage. 
Zheng \textit{et al.}~\cite{Zheng2024} addressed a AAV-based relay system for emergency communications, where a AAV collects data from ground users and uses virtual antenna arrays to transmit the data to multiple remote base stations via collaborative beamforming. Liu \textit{et al.}~\cite{liu2025joint} proposed a AAV-assisted integrated sensing, calculation, and communication (ISCC) system where AAVs detect targets, process information, and transmit results to collection centers. The study maximized the sensing data while minimizing the AoI through joint optimization of AAV parameters under energy constraints, effectively balancing information freshness with sensing performance. 
Likewise, Liu \textit{et al.}~\cite{Liu2025} introduced an innovative island-clustering approach to minimize AoI by jointly optimizing the transmit power of SNs and AAV flight trajectory in wireless-powered communication networks spanning multiple islands. 

\par Moreover, the optimization of communication scheduling has been proven in recent research to be an effective approach for reducing the AoI and energy consumption. For instance, Samir \textit{et al.}~\cite{Samir2022} proposed an online scheduling policy and dynamic AAV altitude control to ensure timely data delivery when direct communication from IoT devices to the BS is not possible, with the goal of minimizing the expected weighted sum AoI. 
Zakeri \textit{et al.}~\cite{Zakeri2024} investigated a multi-source relaying system where independent sources generate status update packets transmitted to a destination via a relay through unreliable communication links. They developed a structure-aware algorithm to optimize the transmission scheduling to minimize AoI while adhering to the transmission capacity and resource constraints. 
Likewise, Qu \textit{et al.}~\cite{Qu2024} addressed the challenge of maintaining information freshness in wireless sensor networks through optimized scheduling. Specifically, they proposed a deadline-constrained and cooperative multi-RIS-assisted framework to minimize AoI while satisfying the deadline and resource constraints. 
In~\cite{Zhao2025}, Zhao \textit{et al.} introduced parallel optimization and primal-dual methods for the SN side. They determined the optimal transmission power of AAV and developed scheduling strategies for the data center on the AAV side. Additionally, they applied a task-associated genetic algorithm to design the visit order of AAV. 
Chen \textit{et al.}~\cite{chen2025average} addressed AoI optimization in wireless-powered networks with directional charging capabilities. The authors developed a framework that optimizes the charging time computation and directional charging schedule, and proved theoretical AoI bounds related to maximum transmitting intervals of the nodes and charging strategies. 
\par However, these studies typically addressed AoI or energy consumption as separate optimization objectives. Furthermore, these works did not explore the joint optimization of AAV trajectory and communication scheduling, especially in multi-AAV scenarios. 

\subsection{Optimization Methods for AoI and Energy Consumption}
\par Various schemes are used to handle the complex optimization problems in the domain of AAV communications and networks. 
For example, Gao \textit{et al.}\cite{Gao2023} addressed an iterative strategy to improve information freshness in wireless sensor networks and optimized AAV flight trajectory by using an improved ant colony optimization (ACO) algorithm. In\cite{Qi2023}, Qi \textit{et al.} introduced a coalition formation-based group-buying auction method for AAV-enabled data collection in sensor networks. The method encouraged sensors to form coalitions and bid for AAV services. This approach employed a trust auction that ensured truthfulness, social welfare, and an efficient parallel algorithm to determine the optimal coalition structure. Likewise, Hoang \textit{et al.}\cite{Hoang2024} analyzed the performance of a multi-antenna AAV-aided non-orthogonal multiple access (NOMA) multi-user pairing system. They derived exact and asymptotic closed-form expressions for block error rate, throughput, and goodput through Gaussian-Chebyshev quadrature and incomplete gamma function approximations.
Zhan \textit{et al.}\cite{Zhan2024} proposed an AoI-operation time tradeoff framework for AAV sensing in multi-cell cellular networks. They introduced a search algorithm and a low-complexity double graph-based algorithm (DGA) for optimal and suboptimal solutions, respectively. 
Reddy \textit{et al.}\cite{Reddy2025} proposed a novel analytical framework for AoI in heterogeneous server environments. They proposed the proactive obsolete packet management (POPMAN) algorithm that dynamically identifies and discards outdated packets in last-come-first-serve (LCFS) queueing systems by using stochastic hybrid systems methodology. 
\par Moreover, DRL has been increasingly recognized for its real-time adaptability in wireless communications. This recognition offers promising opportunities to dynamically optimize the performance of the communication system and the efficiency of resource utilization. For instance, Zheng \textit{et al.}\cite{Zheng2024a} introduced the age-energy efficiency metric to evaluate information timeliness and energy consumption in wireless powered industrial Internet of everything networks. They modeled the optimization problem as a two-stage markov decision process and proposed a dual-layer deep Q-network algorithm that achieves over 25$\%$ improvement compared to benchmarks. 
Li \textit{et al.}\cite{Li2024b} addressed AoI minimization in wireless powered IoT through joint scheduling of AAVs and sensors despite constraints on half-duplex hardware. They proposed a multi-agent DRL method based on factorized value functions for optimal coordination. 
Moreover, Zhang \textit{et al.}\cite{zhang2025aoi} introduced a two-stage optimization approach to minimize AoI in space-air-ground integrated networks. The first stage paired AAVs with IoT device clusters for trajectory optimization by gale-shapley matching algorithm, while the second stage configured network parameters by soft actor-critic. 
Huang \textit{et al.}\cite{Huang2025} developed a novel iterative learning algorithm to minimize AoI in post-disaster AAV operations. The approach combined deep reinforcement learning guidance with strategic route operations to progressively optimize AAV trajectories while respecting energy limitations and environmental obstacles. 
Likewise, Zhang \textit{et al.}\cite{Zhang2025} proposed a hierarchical reinforcement learning framework for optimizing AoI in NOMA-based edge networks, which decomposes the non-convex optimization problem into three distributed agents that separately handle efficiency scheduling, fairness scheduling, and higher-level policy balancing. 
\par However, these studies primarily focused on real-time state information without considering long-term temporal dependencies. Conventional DRL approaches emphasize immediate rewards, which limits their ability to capture complex temporal patterns in dynamic environments. Furthermore, these methods lack effective feature recalibration mechanisms to adjust input feature importance based on context. 

\subsection{Summary}
\par Overall, the differences between our research and previous works can be summarized as follows. First, certain previous studies overlooked the real-time adaptability and coordination between AAVs, which may limit their ability to establish strong and reliable communication links, especially in remote areas with challenging conditions. Our work specifically addresses this gap by proposing a novel system model that employs distributed beamforming to mitigate flight-induced delays. 
Second, most approaches optimize the AAV trajectory and communication scheduling in isolation, neglecting the synergistic potential of joint spatiotemporal resource allocation. To this end, we formulate a joint optimization problem that simultaneously considers both aspects to minimize the AoI and energy consumption. 
Third, conventional DRL-based methods focus on short-term reward maximization without capturing long-term temporal dependencies, which limits their effectiveness in dynamic AAV environments. Our proposed SAC-TLS algorithm overcomes this limitation by incorporating temporal sequence modeling, LNGRU, and SE block to enhance the feature representation and temporal awareness. 


\begin{figure}[t]
    \centerline{\includegraphics[width=0.45\textwidth]{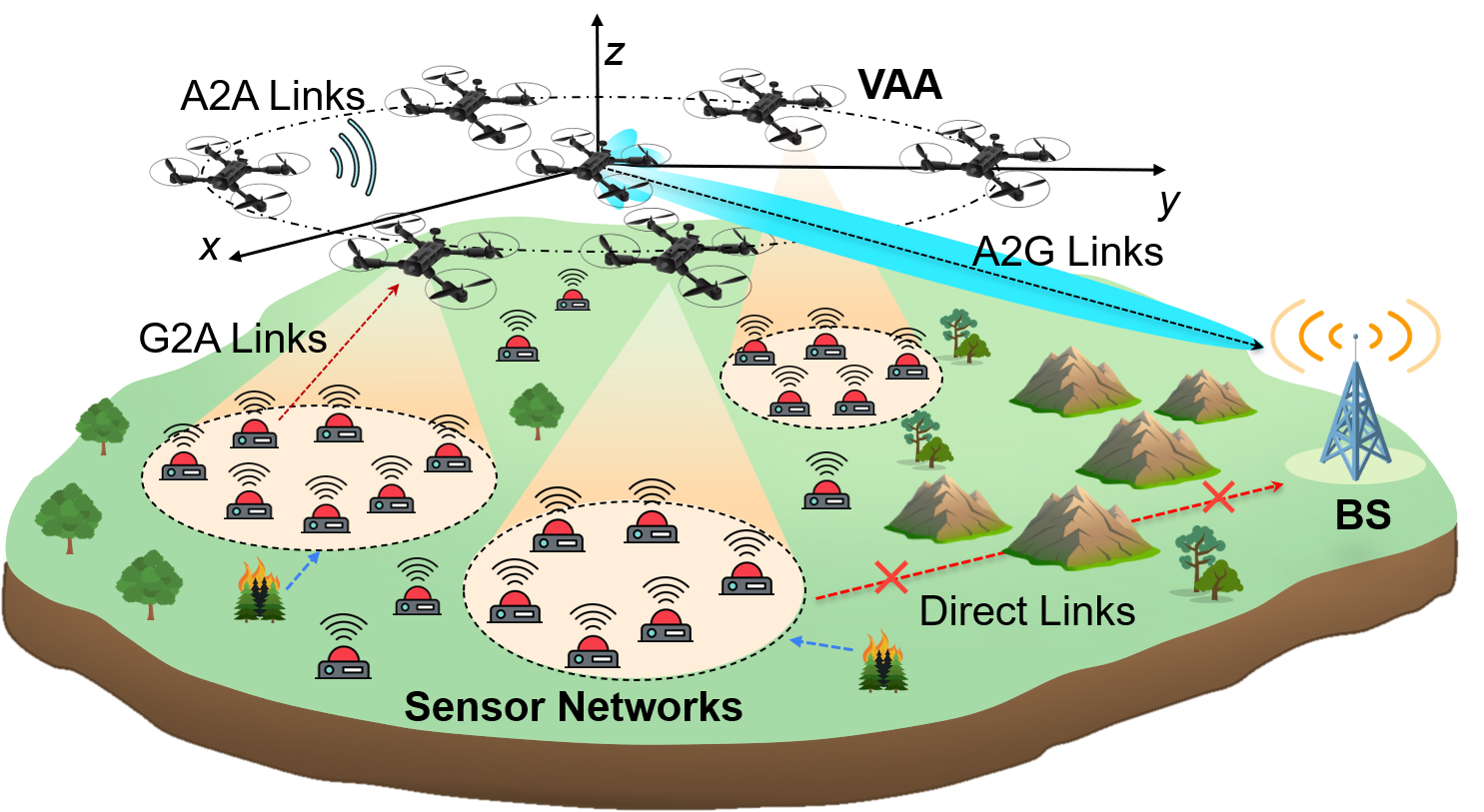}}
    \caption{A AAV-assisted AoI-sensitive data forwarding system in IoT based on distributed beamforming.}
    \label{fig:system model}
 \end{figure}
 \begin{figure}[t]
    \centerline{\includegraphics[width=0.4\textwidth]{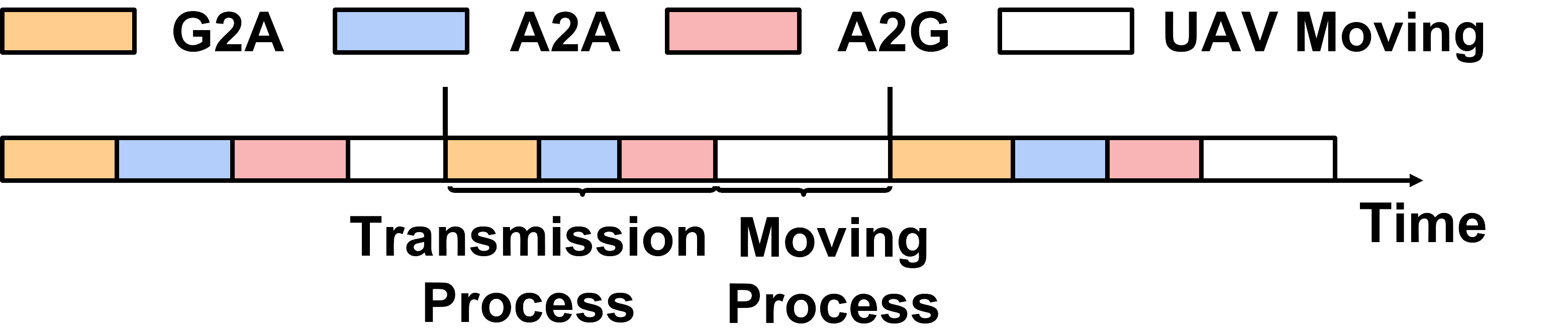}}
    \caption{Time allocation for transmission and moving processes.}
    \label{fig:time slot}
 \end{figure} 
\section{System Models}\label{sec:system model}
\par In this section, we first propose the system overview. Subsequently, we introduce the considered models, which includes the network, AoI, and AAV energy cost models, to characteristize the optimization objectives and decision variables. The main notations used in this paper are summarized in Table~\ref{tab:notations}. 

\begin{table}[t]
\centering
\caption{Main notations}
\label{tab:notations}
\begin{tabular}{ll}
\toprule
\textbf{Notation} & \textbf{Definition} \\ \midrule
$t$, $T$, $\mathcal{T}$ & The index, the number, and the set of time slots \\
$i$, $N_{SN}$, $\mathcal{N}$ & The index, the number, and the set of SNs \\
$j$, $N_{AAV}$, $\mathcal{U}$ & The index, the number, and the set of AAVs \\ 
$\mathcal{D}$ & The sets of the volume of data for SNs \\ 
$\boldsymbol{q}_{i}^{SN}$ & Coordinate of the $i$-th SN \\ 
$\boldsymbol{q}_{j}^{AAV}(t)$ & Coordinate of the $j$-th AAV in time slot $t$ \\ 
$H^U$ & The flight altitude of AAVs \\ 
$a_{j}^{x}(t)$, $a_{j}^{y}(t)$ & Flying distances of AAV in time slot $t$ \\ 
$X^{min}$, $X^{max}$ & Border of target area on the x-axis \\ 
$Y^{min}$, $X^{max}$ & Border of target area on the y-axis \\ 
$d_{min}$ & The minimum separation distance of AAVs \\ 
$v_{max}$& The maximum speed of the AAV \\ 
$U_{tip}$ & Tip speed of the rotor blade \\ 
$v_0$ & Mean rotor induced velocity in hover \\ 
$d_0$ & Drag ratio \\ 
$\rho$ & Air density \\ 
$s$ & The rotor solidity \\
$g_{i,j}$ & Channel gain between the $i$-th SN and $j$-th AAV \\
$h_{j,j'}$ & Channel gain between the $j$-th AAV and $j'$-th AAV \\ 
$g_0$ & Channel gain between the VAA and BS  \\ 
$\sigma^{2}$ & Noise power \\ 
$B$ & Bandwidth \\ 
\bottomrule
\end{tabular}
\end{table}

\subsection{System Overview}
\par As shown in Fig.~\ref{fig:system model}, we consider a AAV-assisted AoI-sensitive data forwarding system in IoT. In this system, the SNs denoted as $\mathcal{N}\triangleq \{i\mid 1,2,\dots,N_{SN}\}$ are randomly distributed across the monitor area. We consider that each SN generates data in each time slots and the volume of data for each SN is denoted by $\mathcal{D}\triangleq \{D_{1},D_{2},\dots,D_{N_{SN}}\}$. Due to the complex terrestrial network environments and long distance between these SNs and BS, the direct transmissions between them may be infeasible. Under the circumstances, a fleet of AAVs denoted as $\mathcal{U}\triangleq \{j\mid 1,2,\dots,N_{AAV}\}$ are deployed to collect and forward the data from the sensor network to the remote BS. Note that each AAV is equipped with a single omnidirectional antenna. In order to reduce the information delay caused by the AAV flying back and forth between the SNs and the BS, we consider introducing distributed beamforming method and constructing the AAVs as a VAA to enhance the transmission ranges without moving AAVs, thereby reducing the AoI of these SNs. 

\par Based on this, we consider a time-slotted muti-access protocol, which is shown in Fig.~\ref{fig:time slot}. Specifically, each frame is divided into multiple time slots with a unit length. As such, the set of all time slots is denoted by $\mathcal{T}\triangleq \{t\mid 1,2,\dots,T\}$. Following this, each time slot is divided into four phases, $\delta_{G2A}(t)$ for G2A transmission, $\delta_{A2A}(t)$ for A2A transmission, $\delta_{A2G}(t)$ for A2G transmission, and $\delta_{move}(t)$ for the AAV moving process. During the G2A transmission phase, each AAV collects sensing data from SNs within its coverage area. In the A2A transmission phase, the AAVs broadcast the collected data to other AAVs. In the A2G transmission phase, the AAVs form a VAA to communicate with the BS using distributed beamforming. Finally, in the AAV moving phase, the AAVs fly to their next designated hovering positions.

\par For generality, we adopt a 3D Cartesian coordinate system, where the position of the $i$-th SN is $\boldsymbol{q}_{i}^{SN} = (x_{i}, y_{i}, 0)$, and the position of the $j$-th AAV in time slot $t$ is $\boldsymbol{q}_{j}^{AAV}(t) = (x_{j}(t), y_{j}(t), H^U)$. Subsequently, we will detail the network model, which contains three key models with respect to transmission, i.e., the G2A, A2A, and A2G transmission models. Following this, we will derive an AoI model and an AAV energy cost model. 

\subsection{Network Model} 
\par In this section, we introduce a detailed introduction of the G2A, A2A, and A2G transmission models.

\par {\textit{1) G2A Transmission Model.}}
In this part, we apply the frequency division multiplexing access (FDMA) mechanism and calculate the transmission rate using probabilistic LoS as the communication link between the SNs and AAVs. Specifically, we consider that different SNs utilize different frequencies to prevent interference during data transmission. Besides, we assume that a SN can only communicate with one AAV at a time, while a AAV can communicate with multiple SNs simultaneously. To this end, we define a binary variable $\beta_{i,j}(t)$ to represent whether the $j$-th AAV communicates with the $i$-th SN in time slot $t$. Therefore, the communication scheduling constraints for SNs are as follows:
\begin{align}
    \sum_{j = 1}^{N_{AAV}} \beta_{i,j}(t)\leq 1,\beta_{i,j}(t)\in \{0,1\},\forall i\in\mathcal{N}, \forall j\in\mathcal{U} . \label{beta}
\end{align}

\par Then, we consider a wireless channel model between the $i$-th SN and the $j$-th AAV, which includes both line-of-sight (LoS) and non-line-of-sight (NLoS) transmission links. The probability of LoS is given by 
\begin{align}
    P_{LoS}(\theta_{i,j}(t))=\frac{1}{1+m \exp(-180 b(\theta_{i,j}(t))/\pi-n)} ,
\end{align}
where $m$ and $n$ are two constants related to the environment, and $\theta_{i,j}(t)=\sin^{-1}(h_{j}^{AAV}(t)/d_{i,j}^{H})$, where $h_{j}^{AAV}(t)$ and $d_{i,j}^{H}$ are the vertical and horizontal distance between the $j$-th AAV and the $i$-th SN, respectively. Correspondingly, the probability of NLoS is $P_{NLoS}=1-P_{LoS}(\theta)$. Furthermore, we support that the conditions of the uplink and downlink channels between SNs and AAVs are comparable. In this case, the average channel power gain between the $i$-th SN and the $j$-th AAV can be expressed as:
\begin{align}
    g_{i,j}=\frac{1}{P_{LoS}L_{LoS}+P_{NLoS}L_{NLoS}}.
\end{align} 
Then, the uplink data transmission rate between the AAV and the SN is given by 
\begin{align}
    R_{i,j}=B_{i,j}\log_{2}(1+\frac{P_{i}g_{i,j}}{\sigma^{2}}),
\end{align} 
where $P_{i}$ is the transmit power and $\sigma^{2}$ is the noise power.

\par By collecting data from accessible SNs, the capacity of the $j$-th AAV during the G2A transmission is given by 
\begin{align}
    S_{j}=\sum_{i=1}^{N_{SN}} \beta_{i,j}D_{i}.\label{su}
\end{align}
\par Then, the transmission time of $j$-th AAV at time slot $t$ is denoted as $\delta_{j}(t)=\sum_{i=1}^{N_{SN}} (\beta_{i,j}D_{i})/R_{i,j}$. Thus, the phase of G2A transmission is $\delta_{G2A}(t)=\max \left( \delta_j(t) \mid 1 \leq j \leq N_{\text{AAV}} \right)$.

\par{\textit{2) A2A Transmission Model.}
Upon collecting complete data from a SN, the AAV receiver will broadcast the data to all AAVs. Here, we consider that the AAV adopts FDMA to broadcast data to other AAVs. Given the high operating altitude of AAVs, a LoS channel model should be used for A2A transmission. We use $d_{j,j'}$ to indicate the distance between the $j'$-th AAV and the $j$-th AAV, and the channel power gain between $j$-th AAV and $j'$-th AAV can be expressed by $h_{j,j'}=\rho _{0}/{d_{j,j'}^{\alpha}}$, where parameter $\rho _{0}$ is the channel power gain associated with a reference distance and parameter $\alpha$ is the path loss exponent. Then, the transmission rate during the aerial broadcast phase is given by
\begin{align}
    R_{j,j'}^{A2A} = B_{j'} \log_{2} \left( 1 + \frac{P_{j'}h_{j,j'}}{\sigma^{2}} \right),
\end{align}
where parameter $B_{j'}$ is the channel bandwidth allocated to $j'$th AAV and $P_{j'}$ is the transmit power of $j'$-th AAV. Due to the characteristics of broadcasting, the actual broadcast rate of the $j$-th AAV should be the minimum rate $R_{j}^{A2A}$ achievable by all AAVs to ensure simultaneous data reception. 

\par Since the distances between AAVs to each other are much smaller than the distances between AAVs and BS, we assume that all data collected from SNs can be reliably broadcast to the AAV swarm through short-range communications within each time slot. In this case, the phase of A2A transmission is given by 
\begin{align}
    \delta_{A2A}(t)=\max\limits_{1\leq j \leq N_{\text{AAV}}} \frac{S_{j}}{{R_{j}^{A2A}}},
\end{align}  
where $S_{j}$ is data successfully collected by the $j$-th AAV which is denoted in Eq. (\ref{su}).

\par\textit{3) A2G Transmission Model Based on AAV-Enabled VAA.}
After the data broadcast to all AAVs, we use distributed beamforming for data transmission between AAVs and the BS. We consider the channel condition between the VAA and the BS is LoS. Then, the signal-to-noise ratio (SNR) of the BS can be expressed as $\gamma_{SNR}(t)=(\sum_{\forall j\in \mathcal{U}} \sqrt{P_{j}(t)g_{0}d_{j,BS}^{-\alpha}})^2/{\sigma^2}$~\cite{Li2024a}, where $P_{j}(t)$ represents the transmit power at time slot $t$, $g_{0}$ represents the channel power gain, $d_{j,BS}$ denotes the propagation distance between $j$-th AAV and BS, $\alpha$ is the path loss exponent. Based on this, the achievable rate from the VAA to the BS is given by
\begin{align}
    R_{BS}(t)=B\log_{2}(1+\gamma_{SNR}(t)),
\end{align}
where $B$ is the carrier bandwidth. Then, the data successfully forwarded from the VAA to the BS in time slot $t$ can be denoted by $S_{A2G}=\delta_{A2G}(t)R_{BS}(t)$, 
where $\delta_{A2G}(t) \leq 1-\delta_{G2A}(t)-\delta_{A2A}(t)-\delta_{move}(t)$. Note that distributed beamforming is to calculate the maximum capacity of the VAA to the BS without optimization. 

\subsection{AoI Model}
\par In this work, data validity is closely associated with the timeliness of its collection and transmission. Thus, we adopt the concept of AoI $A_{i}(t)$ to reflect the freshness of the data from $i$-th SN at time slot $t$. 

\par If the $i$-th SN fails to communicate with any AAV in time slot $t$, its data incurs an additional delay, and the AoI updates as $A_{i}(t+1) = A_{i}(t) + 1$. If the $j$-th AAV broadcasts to all AAVs and forwards the data from the $i$-th SN to the BS, the AoI of the $i$-th SN will decrease in the following time slot.

\par Due to limited channel capacity, the VAA may not always successfully forward data to the BS during the A2G phase. In this case, the fraction of data successfully forwarded can be expressed as: 
\begin{align}
    Q(t)\triangleq \frac{\min\{S_{G2A},S_{A2G}\}}{S_{c}},
\end{align} 
where $S_{c}=\sum_{i=1}^{N_{SN}} D_{i}$ is the size of sensing data of all SNs before data collection
and $S_{G2A}=\sum_{j=1}^{N_{AAV}} S_{j}$ is the size of received data during G2A transmission, where $S_{j}$ is shown in Eq. \eqref{su}. Thus, the AoI dynamics of SNs can be summarized as follows: 
\begin{align}
    A_{i}(t+1)=
    \begin{cases}
        (1-Q(t))(A_{i}(t)+1), \text{if}\enspace\beta_{i,j}(t)=1,\\
        A_{i}(t)+1,\quad\text{otherwise.}\label{con:inventoryflow} 
    \end{cases}
\end{align} 

\subsection{AAV Energy Cost Model}
\par In this part, we detail the following energy consumption model for AAVs. In particular, the dominant component of AAV energy consumption is propulsion energy consumption, whereas the energy consumed for communication is comparatively negligible, as it is approximately two orders of magnitude smaller than the propulsion energy. Thus, for an altitude-fixed rotary-wing AAV, we denote the propulsion power as $P_{AAV}(v)$, which can be expressed as:
\begin{align}
    P_{AAV}(v) = P_0 \left(1 + \frac{3 v^2}{U_{tip}^2}\right) + P_1 \sqrt{1 + \frac{v^4}{4 v_0^4} - \frac{v^2}{2 v_0^2}} \nonumber\\
+ \frac{1}{2} \rho d_0 s A v^3,
\end{align}
where $P_0$ represents the power associated with blade profile drag, and $P_1$ denotes the derived power during required for AAV to maintain hover, $U_{tip}$ corresponds the tip velocity of the rotor blade, $v_0$ indicates the average induced velocity during hovering, $d_0$ and $s$ are the fuselage drag coefficient and rotor solidity, respectively, and $\rho$ denotes the ambient air density. 
\par Thus, the flight energy consumption of $j$-th AAV is given by 
\begin{align}
    E_{j}^{move}=\sum_{t=1}^{T} P_{AAV}(v)\delta_{move}(t).
\end{align} 
The hovering energy consumption can be expressed as: 
\begin{align}
    E^{hov}_{j}=\sum_{t=1}^{T} P_{hov}(\delta_{G2A}(t)+\delta_{A2A}(t)+\delta_{A2G}(t)),
\end{align} 
where $P_{hov}=P_0+P_1$. Then, the energy consumption of the $j$-th AAV can be denoted as 
\begin{align}
    E_j=&E_{j}^{move}+E^{hov}_{j}.
    \label{energy}
\end{align}
Note that we omit the extra energy consumption caused by the acceleration and deceleration of the AAV during horizontal flight, since it constitutes only a minor part of the total operation time in the maneuver duration of the AAV. 

\section{Problem Formulation and Analyses}\label{sec:problem formulation}
\par In this section, we propose an AoI-enhanced and energy-efficient 
AAV-assisted forwarding problem. First, we state the main idea of the optimization problem. Then, the key decision variables and optimization objectives are presented. Finally, we formulate an optimization problem and give the corresponding analysis. 

\subsection{Problem Statement}
\par The considered system concerns two goals, $i.e.$, reducing the time averaged AoI of sensors and energy cost of AAVs. 
\par As shown in Eq. \eqref{con:inventoryflow}, the access control of SNs and mobility control of AAVs in each time slot not only depend on the current system states but also affect the future evolution of the AoI statuses for SNs. Therefore, the design goal to minimize the long-term time-average AoI of all SNs can be achieved by the access control strategy for SNs and the position of AAVs at each time slot. To reduce the AoI of sensors, the AAVs need to fly to better locations and communicate with the SNs and BS, which will result in extra energy costs. Thus, the position changes of AAVs should be minimized by considering the energy efficiency. 

\par We define these decision variables as: $(i)$ $\mathbf{\Phi} \triangleq \{\beta_{i,j}(t)\mid \forall i\in \mathcal{N},\forall j\in \mathcal{U},\forall t\in \mathcal{T}\}$, a binary variable denoted the access control of SNs. $(ii)$ $\boldsymbol{q}\triangleq\{(x_{j}^{AAV},y_{j}^{AAV},H^U)\mid j\in \mathcal{U}\} $, a matrix consisting of continuous variables denoted the 3D hover locations of the AAVs for communicating with different SNs. Based on the aforementioned analysis, the optimization objectives can be presented as follows. 
\par\textit{ Optimization Objectives 1:} Our first optimization objective is to minimize the time averaged AoI of sensors to enhance the 
information freshness in sensor networks. To this end, we need to jointly optimize the $\mathbf{\Phi}$ and $\boldsymbol{q}$. Thus, the first optimization objective is given by 
\begin{align}
    f_{1}(\mathbf{\Phi},\boldsymbol{q}) =\frac{1}{T} \frac{1}{N} \sum_{t=1}^{T} \sum_{i=1}^{N_{SN}} A_i(t).\label{aoi}
\end{align}

\par\textit{Optimization Objectives 2:} Our second optimization objective is to minimize the energy costs of all AAVs to adjust their positions. 
We optimize $\boldsymbol{q}$ to achieve this goal, and the second optimization objective 
is given by
\begin{align}
    f_{2}(\boldsymbol{q}) =\sum\nolimits_{j\in \mathcal{U}} E_{j}.
\end{align}

\par By considering these conflicting objectives, the optimization problem can be expressed as follows:  
\begin{subequations}\label{opti}
\begin{align}
(P1) : \mathop{\min}\limits_{\mathbf{\Phi},\mathbf{Q}} 
 & \ F= \{f_1,f_2\}, \\
\text{ s.t. }
    &X^{min} \leq x_{j}^{AAV} \leq X^{max}, \label{x_AAV}\\ 
    &Y^{min} \leq y_{j}^{AAV} \leq Y^{max}, \label{y_AAV}\\ 
    &d_{j1,j2}\geqslant d_{min},\label{d_min}\\
    &\|\boldsymbol{q}_{j}^{AAV}(t+1)-\boldsymbol{q}_{j}^{AAV}(t)\|\leq v_{max}\delta_{move}(t), \label{v_max}
\end{align}
\end{subequations}
where Eqs. \eqref{x_AAV} and \eqref{y_AAV} represent the allowable flight regions for the AAV, and any operation beyond these regions is regarded as a boundary constraint violation. Moreover, Eqs. \eqref{d_min} and \eqref{v_max} show the constraints of the flight of AAVs. 

\subsection{Problem Analyses}
\par The problem $(P1)$ exhibits the following characteristics. First, the problem $(P1)$ is inherently non-convex~\cite{Long2022}, mainly because its initial objective function involves interdependent variables, including both continuous variables $\boldsymbol{q}$ and discrete decision variables $\mathbf{\Phi}$. 
Second, the problem $(P1)$ incorporates long-term optimization objectives that are impacted by the trajectory of AAVs and the current availability status of SNs. Finally, the problem $(P1)$ is a formulation with conflicting optimization objectives. For example, to minimize AoI, the AAV must frequently accelerate or hover to establish communication with the SNs. However, this behavior leads to an increase in energy consumption. Likewise, energy efficiency necessitates restricting the movement of AAVs. This limitation may result in accumulation of AoI due to communication delays. 

\par Therefore, the problem $(P1)$ constitutes a non-convex mixed-integer programming formulation with long-term and dynamic optimization objectives. Due to the inherent complexity of $P1$, traditional offline methods, such as convex optimization and evolutionary computing, are generally unsuitable for solving this problem. 
Moreover, the problem $(P1)$ is characterized by conflicting optimization objectives, whose relative importance may vary in specific practical scenarios and application contexts. 
This dynamic trade-off underscores the need for adaptive strategies that balance competing goals based on real-time conditions. Additionally, environmental uncertainties ($e.g.$, stochastic channel conditions, unpredictable data generation rates) further complicate the optimization problem. 
\par In this case, DRL emerges as a promising framework, which can inherently learn dynamic policies through online interactions, capturing temporal dependencies. The aforementioned reasons motivate us to propose a DRL approach for solving the formulated problem. 

\section{The Proposed SAC-TLS}\label{sec:SAC-TLS}
\par This section presents a DRL-based method to address the formulated optimization problem. In pursuit of this objective, we first discuss the motivations for using DRL and reformulate our optimization problem as a markov decision process (MDP). Then, we detail the proposed SAC-TLS algorithm with several enhancements. 

\subsection{Motivations and MDP}
\par In this part, we first introduce the motivations for using DRL, and then present the formulation and simplification of the MDP. 
\par\textit{1) Motivations for Using DRL.}
The considered problem exhibits significant dynamic and uncertain properties, with factors like AAV mobility and changes over multiple transmission phases. Traditional optimization approaches, including both convex and non-convex methods, are often inadequate for handling such time-varying environments~\cite{Li2024,Wang2023}. In contrast, DRL provides the capability to adaptively learn robust policies in complex and uncertain scenarios. Consequently, we adopt a DRL-based framework to address the underlying optimization problem. 

\par\textit{2) MDP Formulation.}
We reformulate the optimization problem defined in Eq. \eqref{opti} as an MDP. Mathematically, an MDP is formally represented by the tuple \( (S, A, P, R, \gamma) \), where $S$ is the state space, $A$ is the action space, $P$ denotes the probability of state transition, $R$ represents the reward function, and $\gamma$ is the discount factor. Among these, the definitions of state, action, and reward are of primary importance and are elaborated as follows. 
\begin{itemize}
    \item \textbf{State Space:} The state space is designed to capture essential spatial and operational dynamics influencing system performance. Specifically, the positions of both AAVs and SNs, along with the AoI of each SN, are incorporated to reflect spatial dynamics. Thus, the state at time slot $t$ is formally expressed as: 
    \begin{align}
        s_t = \{ \boldsymbol{q}^{AAV}_t, \boldsymbol{q}^{SN}_t,\boldsymbol{A}^{SN}_t\}.
    \end{align}
    \item \textbf{Action Space:} In our system, the AAV can adjust its position to optimize its communication with SNs. Additionally, the binary variable $\beta_{i,j}(t)$ reflects the communication scheduling decision, which determines whether the $j$th AAV communicates with the $i$th SN in time slot $t$. As such, the action of agent at time slot $t$ is represented as: 
    \begin{align}
        a_j(t) = \{a_j^{x}(t), a_j^{y}(t)\},
    \end{align}
    where $a_j^{x}(t)$ and $a_j^{y}(t)$ denote the movement of $j$-th AAV in the $x$ and $y$ directions at time slot $t$, respectively.
    \item \textbf{Reward Function:} The reward is central to effective decision-making, as it directly influences the ability of agents to solve the underlying task. In our framework, it reflects both the optimization objective and the associated constraints, as formulated in Eq. \eqref{reward}. Specifically, the reward assigned to the $j$-th agent is defined as: 
\begin{align}
    r_j(t)= -\rho_1A-\rho_2E+\rho_3c_j-p_j, \label{reward}
\end{align}
where $\rho_1$, $\rho_2$, $\rho_3$ are three normalization parameters to adjust to bring these terms to the same order of magnitude, $c_j$ is the amount of SNs covered by the $j$-th AAV to guide agents to avoid the impact of extreme negative rewards, and $p_j$ is the out-of-bounds penalty.

\end{itemize}  
\par\textit{3) MDP Simplification.}
To efficiently handle the complexity arising from the mixed discrete and continuous action space, we simplified the original MDP. Specifically, the discrete nature of communication actions introduces considerable randomness into the decision making process. This randomness complicates the decision making and learning process, and leads to inefficiencies. As a consequence, the selected SNs may fall outside the communication range of AAVs, while SNs within the radio range remain unselected, which leads to wasted communication resources. 
\par To resolve these issues, we introduce dynamic proximity-based action mapping (DPAM), a strategy that simplifies decision-making by directly linking AAV trajectories to communication actions~\cite{Sun2024a}. In particular, DPAM makes communication decisions deterministically by checking whether SNs are within the communication radius of the AAV, thereby eliminating the unpredictability associated with binary actions. This reformulation ensures reliable data collection, enhances energy efficiency, and streamlines the action space. 

\subsection{Standard SAC Algorithm}

\par In this paper, we adopt the soft actor-critic (SAC) algorithm as our optimization framework~\cite{Haarnoja2018}. SAC is a model-free, off-policy reinforcement learning method that is particularly suited for high-dimensional and continuous action spaces. The primary objective of SAC is to maximize both the expected cumulative reward and the entropy of the policy. By incorporating an entropy term into the objective function, SAC not only seeks to optimize the reward but also encourages exploration, which ensures the agent avoids prematurely converging to a deterministic policy. This formulation can be expressed as:
\begin{align}
    \max_{\theta} J(\theta) = \mathbb{E}_{\pi_\theta} \left[ \sum\nolimits_{t=1}^{T} \gamma^t (r_t + \alpha \mathcal{H}(\pi(\cdot|s_t))) \right],
\end{align}
where \( r_t \) is the reward at time step \( t \), \( \gamma \) is the discount factor, and \( \mathcal{H}(\pi(\cdot|s_t)) \) represents the entropy of the policy \( \pi_\theta \) at state \( s_t \), which is given by 
\begin{align}
    \mathcal{H}(\pi_\theta) = -\mathbb{E}_{a_t \sim \pi_\theta} \left[ \log \pi_\theta(a_t | s_t) \right].
\end{align}
Additionally, the temperature parameter \( \alpha \) plays a crucial role in controlling the balance between reward maximization and entropy regularization and encourages exploration. The temperature \( \alpha \) can be expressed as:
\begin{align}
\alpha = \frac{1}{|\mathcal{A}|} \sum_{a \in \mathcal{A}} \log \pi_\theta(a | s_t),
\end{align}

\par SAC is composed of three main components, which include an actor network \( \pi_\theta \) that generates a stochastic policy, and two critic networks \( Q_\phi \), used to estimate the action-value function \( Q(s_t, a_t) \), representing the expected return for taking action \( a_t \) at state \( s_t \). The critic networks play a crucial role in stabilizing learning by minimizing the Bellman error. This error is given by the loss function as follow:
\begin{align}
L_Q(\phi) = \mathbb{E} \Big[ \Big( Q_\phi(s_t, a_t) - \Big( r_t + \gamma \Big( Q_{\bar{\phi}}(s_{t+1}, a_{t+1}) \nonumber\\- \alpha \log \pi_\theta(a_{t+1} | s_{t+1}) \Big)\Big)\Big)^2 \Big],
\end{align}
where the term \( Q_{\bar{\phi}}(s_{t+1}, a_{t+1}) \) is the target Q-value from the target critic network, and the expression \( \log \pi_\theta(a_{t+1} | s_{t+1}) \) is the log-probability of the next action under the current policy. The inclusion of the entropy term in the Bellman backup helps prevent overestimation of the Q-values and ensures sufficient exploration of the state space.

\par To update the policy, SAC maximizes the following objective, which is based on the value function produced by the critics:
\begin{align}
    J_{\text{policy}}(\theta) = \mathbb{E}_{s_t \sim \mathcal{D}} \left[ \mathbb{E}_{a_t \sim \pi_\theta} \left[ \alpha \log \pi_\theta(a_t | s_t) - Q_\phi(s_t, a_t) \right] \right],
\end{align}
where \( Q_\phi(s_t, a_t) \) represents the estimated Q-value for taking action \( a_t \) at state \( s_t \), and \( \alpha \log \pi_\theta(a_t | s_t) \) is the entropy term, which is crucial as it encourages the policy to explore diverse actions to prevent the policy from becoming overly deterministic and enhance the stability of the learning process. 

\par SAC also benefits from its off-policy nature, where it learns from a replay buffer that stores past experiences. By sampling from this buffer, SAC can learn from data generated by different policies, which improves sample efficiency compared to on-policy algorithms. This off-policy learning makes SAC suitable for environments where interaction with the system is costly or time-consuming, as it can learn from past data instead of requiring fresh interactions at each step.

\par The combination of these components, actor and critic networks, entropy regularization, and off-policy learning, makes SAC a powerful algorithm, capable of efficiently solving reinforcement learning tasks in complex, high-dimensional continuous action spaces. The ability of SAC to balance exploration and exploitation through entropy maximization, along with its stability and efficiency, makes it a suitable and ideal choice for a wide range of reinforcement learning problems.

\par However, in our MDP, the reward function and state are highly variable, which poses challenges for the standard SAC algorithm in terms of the rapid adaptation of its policy, which ultimately leads to slow convergence. Therefore, we propose several enhancements to SAC tailored to our scenario.
\subsection{SAC-TLS Algorithm} 
\par In this work, we introduce an enhanced SAC-TLS. Specifically, SAC-TLS integrates three essential enhancements which are temporal sequence input processing, LNGRU, and SE block, and they are detailed as follows.
\par\textit{1) Temporal Sequence Input Processing.} To improve the learning capability of conventional SAC algorithm in a time-varying environment, we introduce the temporal sequence-based state input. Unlike conventional SAC, which processes each state individually, temporal sequence input allows the algorithm to leverage temporal information from multiple time steps. Specifically, the environment generates a sequence of states $\{s_t, s_{t-1}, ..., s_{t-n}\}$ as the input to actor and critic networks, which enables the agents to better capture dependencies and patterns in the environment to improve policy performance ultimately. 
\par\textit{2) LNGRU.}
To capture long-term dependencies in sequential data, we integrate LNGRU cells into the actor network~\cite{LeiBa2016}. Unlike standard GRU, LNGRU applies layer normalization (LN) after the reset gate \( r_t \), update gate \( z_t \), and candidate hidden state \( \tilde{h}_t \) to stabilize gradient flow during training. This modification addresses the issues of vanishing and exploding gradients, which are common in deep recurrent networks when learning long-term dependencies. In LNGRU, layer normalization is applied to stabilize each gate and the candidate hidden state. Specifically, the normalized gates and candidate hidden state are calculated as:
\begin{align}
    \tilde{h}_t = \tanh\left(LN\left(W_h x_t\right) + r_t \circ LN\left(U_h h_{t-1}\right)\right),
\end{align}
where $LN$  denotes layer normalization, which normalizes the input across each feature dimension to have zero mean and unit variance. This process can be mathematically expressed as: 
\begin{align}
    LN(z) = \frac{z - \mu_z}{\sigma_z + \epsilon},
\end{align}
where $\mu_z$ is the mean of the input $z$, $\sigma_z$ is the standard deviation, and $\epsilon$ is a small constant added for numerical stability. Following this, the normalization mitigates gradient vanishing and explosion, and ensures efficient backpropagation over time. Thus, LNGRU enhances the ability of the algorithm to learn temporal dependencies and improves decision making in dynamic environments. 

\par\textit{3) SE block.}
To enhance the focus of algorithm on critical elements within input sequences, we integrate the SE block~\cite{hu2018squeeze}. Specifically, SE block primarily consists of two key operations, squeeze and excitation, to emphasize key features at each time step. First, the squeeze operation applies global average pooling to the feature map \( h_t \), summarizing each channel into a global descriptor:
\begin{align}
    z_c = \frac{1}{H \times W} \sum_{h=1}^H \sum_{w=1}^W h_t(h, w),
\end{align}
where \( H \) and \( W \) represent the height and width of the feature map, respectively, and \( z_c \) is the global representation for each channel. 
Then, the excitation operation step generates channel attention weights \( \hat{z}_c \) by passing \( z_c \) through fully connected layers and a sigmoid activation: 
\begin{align}
    \hat{z}_c = \sigma(W_2 \delta(W_1 z_c + b_1) + b_2). 
\end{align}
\par We apply SE block to the hidden state \( h_t \) produced by the actor network. After obtaining \( h_t \), the squeeze and excitation operations refine it, enabling the agent to focus on more informative features for action selection. This adjustment improves the learning efficiency by emphasizing critical features while suppressing less important ones. 

\begin{figure*}[t]
    \centerline{\includegraphics[width=0.9\textwidth]{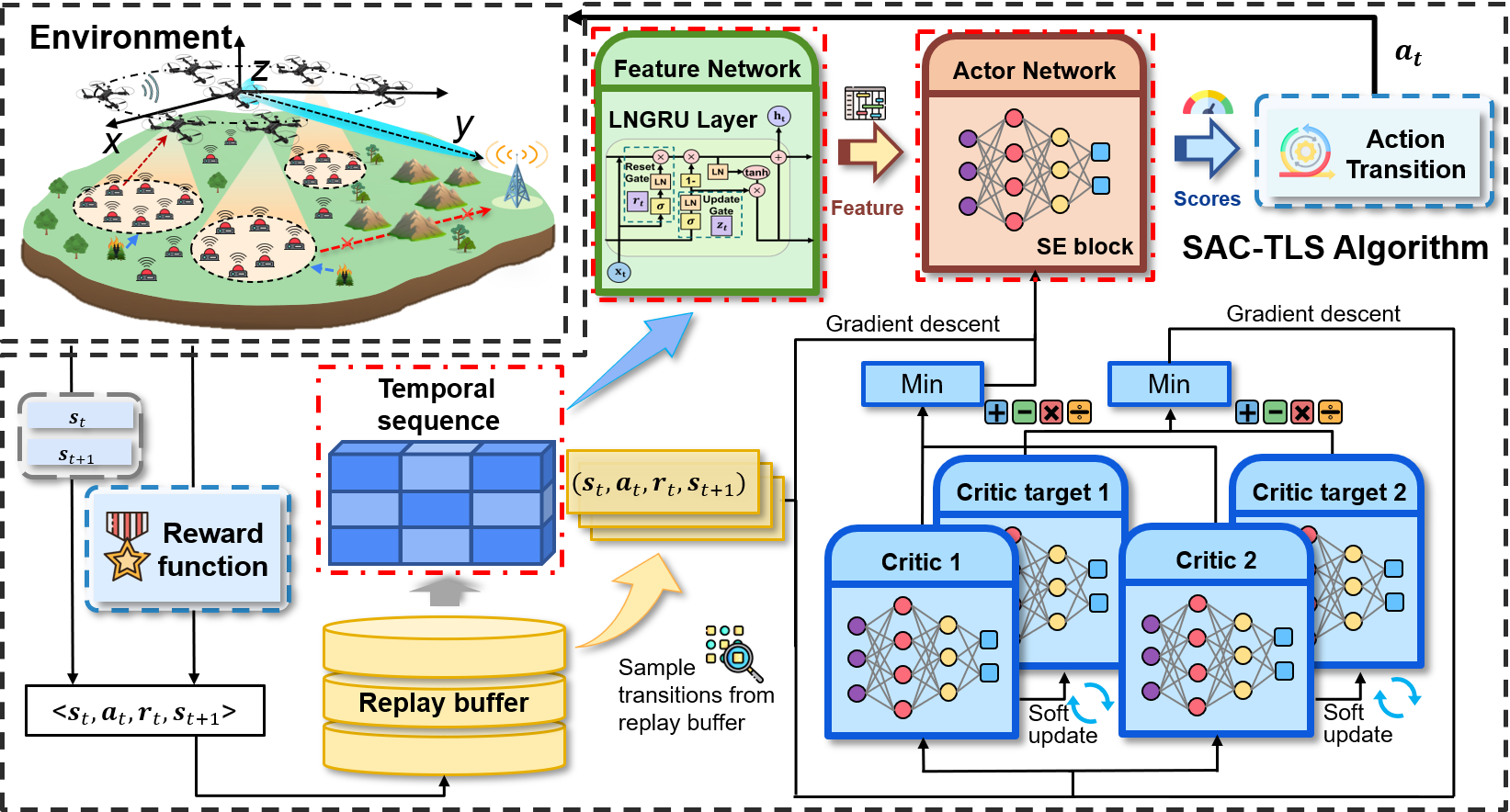}}
    \caption{The framework of the SAC-TLS algorithm.}
    \label{fig:SAC-TLS}
\end{figure*} 
\begin{algorithm}
\caption{SAC-TLS Algorithm}\label{alg:lasac}
    Initialize actor network parameters $\theta$ and critic network parameters $\phi$ \;
    Initialize target critic network $\phi_{\text{target}} \gets \phi$\;
    Initialize SE block parameters $\{W_{\text{SE}}, b_{\text{SE}}\}$\;
    Initialize entropy coefficient $\alpha$ and log $\alpha$\;
    Initialize replay buffer $\mathcal{D}$ \;

    \For{each episode}
    {
        Reset environment and initialize state $s_0$\;
        \For{each time slot $t = 0, 1, ..., T-1$}
        {
            Obtain state sequence $\{s_{t-n}, ..., s_{t-1}, s_t\}$\;
            Compute hidden state $h_t$ using LNGRU Cell from $\{s_{t-n}, ..., s_t\}$\;
            
            Squeeze: $z_t = \text{GlobalAvgPool}(h_t)$ \;
            Excitation: $\hat{h}_t = h_t \odot \sigma(W_{\text{SE}} \cdot z_t + b_{\text{SE}})$\;
            
            Select action $a_t \sim \pi_{\theta}(\cdot|\hat{h}_t)$ \;
            Execute action $a_t$ in environment\;
            Observe next state $s_{t+1}$\;
            Calculate AoI $A$ and energy consumption $E$\;
            Compute reward $r_t$ according to Eq. (\ref{reward})\;
            Store transition $(s_t, a_t, r_t, s_{t+1})$ in replay buffer $\mathcal{D}$\;
        }
        \If{$|\mathcal{D}| \geq$ batch\_size}
        {
            \For{each update step}
            {
                Sample a batch of transitions $\{s, a, r, s'\}$ \;
                Compute target $Q$ values and update critic network\;
                Update actor network\;
                Soft update target network: 
                $\phi_{\text{target}} \gets \tau \phi + (1 - \tau) \phi_{\text{target}}$\;
            }
        }
    }
\end{algorithm}

\par\textit{4) SAC-TLS Framework.} As shown in Fig.~\ref{fig:SAC-TLS}, SAC-TLS extends the core structure of SAC by approximating the policy and the value function through actor and critic networks, respectively. Specifically, the actor network combines LNGRU and SE block to generate action \(a_t\) based on the state sequence. Moreover, the actor network combines LNGRU and the SE block to generate an action \(a_t\) based on the state sequence. Then, the critic network estimates the Q-value of the state-action pair. Each interaction transfer \( (s_t, a_t, r_t, s_{t+1})\) is stored in a replay buffer from which a small batch of samples is drawn to update the network during training. In addition, the main steps of the proposed SAC-TLS algorithm are shown in Algorithm~\ref{alg:lasac}.

\subsection{Complexity Analysis}
\par In this section, we evaluate the computational and space complexity of SAC-TLS during training and execution phases. 
\subsubsection{Training Phase} 
The computational complexity of SAC-TLS in the training phase is formulated as  $O\left(|\theta| + |\phi| + N_{\text{eps}} T |\theta| + \frac{N_{\text{eps}} T K (|\theta| + |\phi|)}{N_b}\right)$, which is summarized as follows: 
\begin{itemize}
    \item \textit{Network Initialization}: This phase focuses on the initialization of network parameters. Specifically, this computational complexity is denoted as $O(|\theta_A| + 2|\phi_C|)$, where $|\theta_A|$ and $|\phi_C|$ denote the number of parameters in the actor network and critic network, respectively.
    \item \textit{Action Sampling}: In this phase, actions are generated via the actor network at each time slot. The complexity is $O(N_{eps} T |\theta_A|)$, where $N_{eps}$ is the number of episodes and $T$ is time slots per episode. 
    \item \textit{Network Update}: In the updating phase, whenever the size of the replay buffer equals the batch size $N_b$, the network parameters will be updated $K$ times. Thus, the complexity for this phase is calculated as
    $O\left(\frac{N_{eps} T K (|\theta_A| + 2|\phi_C|)}{N_b}\right)$. 
\end{itemize}
\par In the training phase, the space complexity of SAC-TLS is determined by the size of the neural network parameters and the replay buffer, which is $O\left(|\theta_A| + 2|\phi_C| + D(2|s| + |a| + 2)\right)$, where $D$ represents the size of the replay buffer and $|s|$, $|a|$ denote the dimensions of the state and action spaces, respectively. 

\subsubsection{Execution Phase}
In the execution phase, the computational complexity of the SAC-TLS algorithm is $O(|\theta_A|)$, as only the parameters of the actor network are stored and used for action selection. Since the critic network and replay buffer are not involved in the execution phase, the space and computational overhead are minimized, which makes the process more efficient during inference.

\section{Simulation Results and Analysis}\label{sec:simulation}
\par In this section, we detail the simulation results and analyses. First, we introduce the simulation setting and benchmarks. Then we present the corresponding results.
\begin{table}[t]
\centering
\caption{Simulation Parameters.}
\label{tab:simulation}
\begin{tabular}{@{}llll@{}}
\toprule
\textbf{Parameter} & \textbf{Value} & \textbf{Parameter} & \textbf{Value} \\
\midrule
$X^{min}$          & 0 m            & $X^{max}$          & 400 m          \\ 
$Y^{min}$          & 0 m            & $Y^{max}$          & 400 m          \\ 
$H^U$              & 15 m             & $T$          & 100             \\ 
$P_0$             & 199.4 W         & $P_1$             & 88.66 W        \\ 
$U_{tip}$           & 120 m/s         & $v_0$             & 4.03           \\
$d_0$               & 0.6             & $\rho$            & 1.225 kg/m³    \\ 
$s$                 & 0.05            & $d_{min}$           & 1 m        \\ 
$a$                &5.18     & $b$  & 0.43 \\ 
$\sigma^2$        & -174 dBm/Hz    & $N_{\text{eps}}$           & 4500    \\
$\gamma$          & 0.99              & $\epsilon$        & 0.02    \\ 
\bottomrule
\end{tabular}
\end{table}
\subsection{Simulation Setups}
\par We consider a typical remote and open environment with absence of significant obstructions. The AAVs fly within the airspace above a square region measuring 400 m $\times$ 400 m. Following this, 60 SNs are divided into clusters, with higher-density clusters concentrated in the central region of the area, while lower-density clusters are distributed toward the edges. Our coordinate system places the left corner of the sensing area as the origin, with the BS coordinates at (2000, 2000, 0), and the initial coordinates of four AAVs at (0, 0, 15), (0, 400, 15), (400, 0, 15), (400, 400, 15), respectively. Other parameters can be found in Table~\ref{tab:simulation}.  
\par For comparison, we utilize the greedy policy, along with twin delayed deep deterministic policy gradient (TD3)~\cite{Fujimoto2018}, proximal policy optimization (PPO)~\cite{Schulman2017}, truncated quantile critics (TQC)~\cite{Kuznetsov2020} and soft actor-critic (SAC)~\cite{Haarnoja2018} as benchmark methods. 
\begin{figure*}[t]
    \centerline{\includegraphics[width=\linewidth]{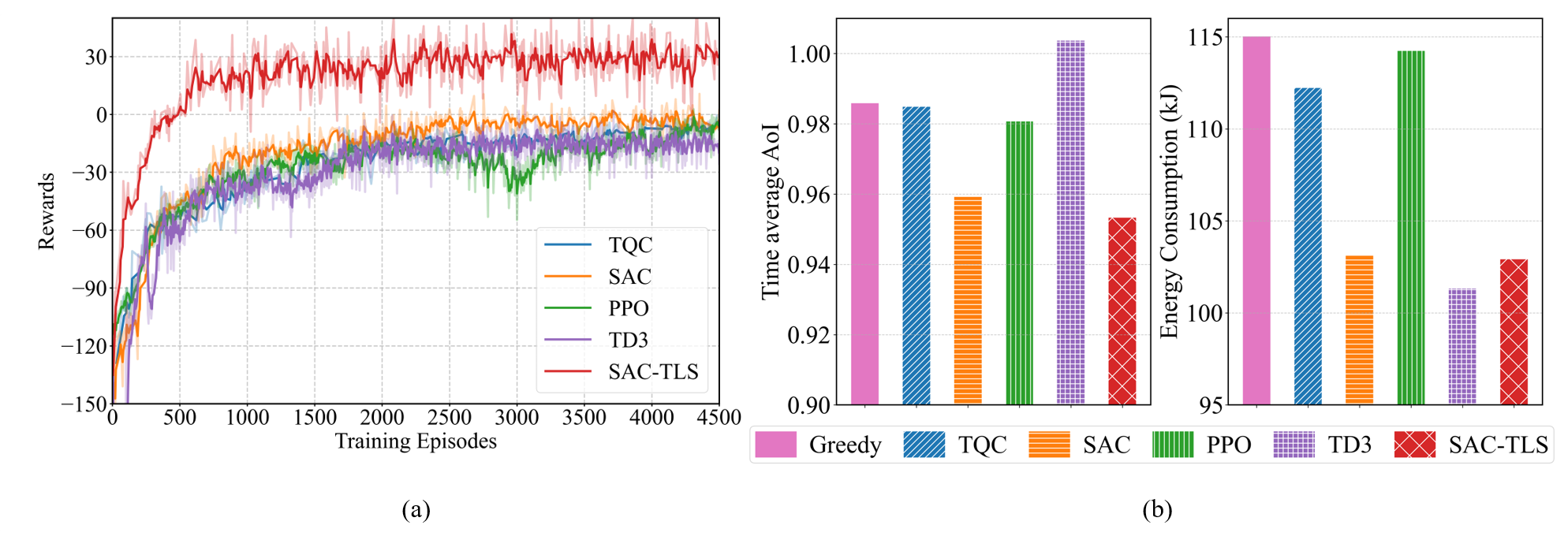}}
    \caption{Simulation results. (a) Cumulative rewards training curve. (b) The optimization objective values of greedy, TQC, SAC, PPO, TD3, and SAC-TLS.}
    \label{fig:Simulation results}
\end{figure*}
\begin{itemize}
    \item \textit{Greedy policy:}
    The greedy algorithm is a simple, heuristic approach that makes locally optimal decisions at each step with the aim of finding the global optimum. Although this approach is computationally efficient and straightforward, it can be limited by its lack of exploration, potentially leading to suboptimal solutions in environments with complex or dynamic state spaces. 
    \item \textit{TD3:}
     TD3 is an improved version of deep deterministic policy gradient (DDPG)~\cite{Lillicrap2016} that addresses overestimation bias by using two Q-networks and enhances stability with delayed updates. It also incorporates target smoothing to reduce variance in Q-value estimates. TD3 is particularly effective in high-noise and uncertain environments, which improves training stability and the accuracy of Q-value predictions.
    \item \textit{PPO:}
    PPO is an on-policy reinforcement learning algorithm that optimizes a surrogate objective function using a clipped probability ratio. This approach limits the policy update within a specified range, preventing large, destabilizing updates. PPO alternates between sampling data and updating the policy with multiple epochs, offering a balance between the stability of value-based methods and the efficiency of policy-gradient approaches.
    \item \textit{TQC:}
    TQC is an off-policy deep reinforcement learning algorithm that addresses Q-value overestimation in methods like DDPG and TD3. It uses quantile regression to capture the distribution of Q-values and truncates extreme quantiles to improve stability and reduce the impact of outliers. This approach provides a more accurate estimate of the Q-value distribution, enhancing the reliability of value-based methods in continuous action spaces.
    \item \textit{SAC:}
    SAC is an off-policy algorithm that maximizes both reward and entropy, encouraging exploration through entropy regularization. It uses two Q-functions and a stochastic policy, improving stability and performance in continuous action spaces by balancing exploration and exploitation.
\end{itemize}
\begin{figure}
    \centering
    \includegraphics[width=0.9\linewidth]{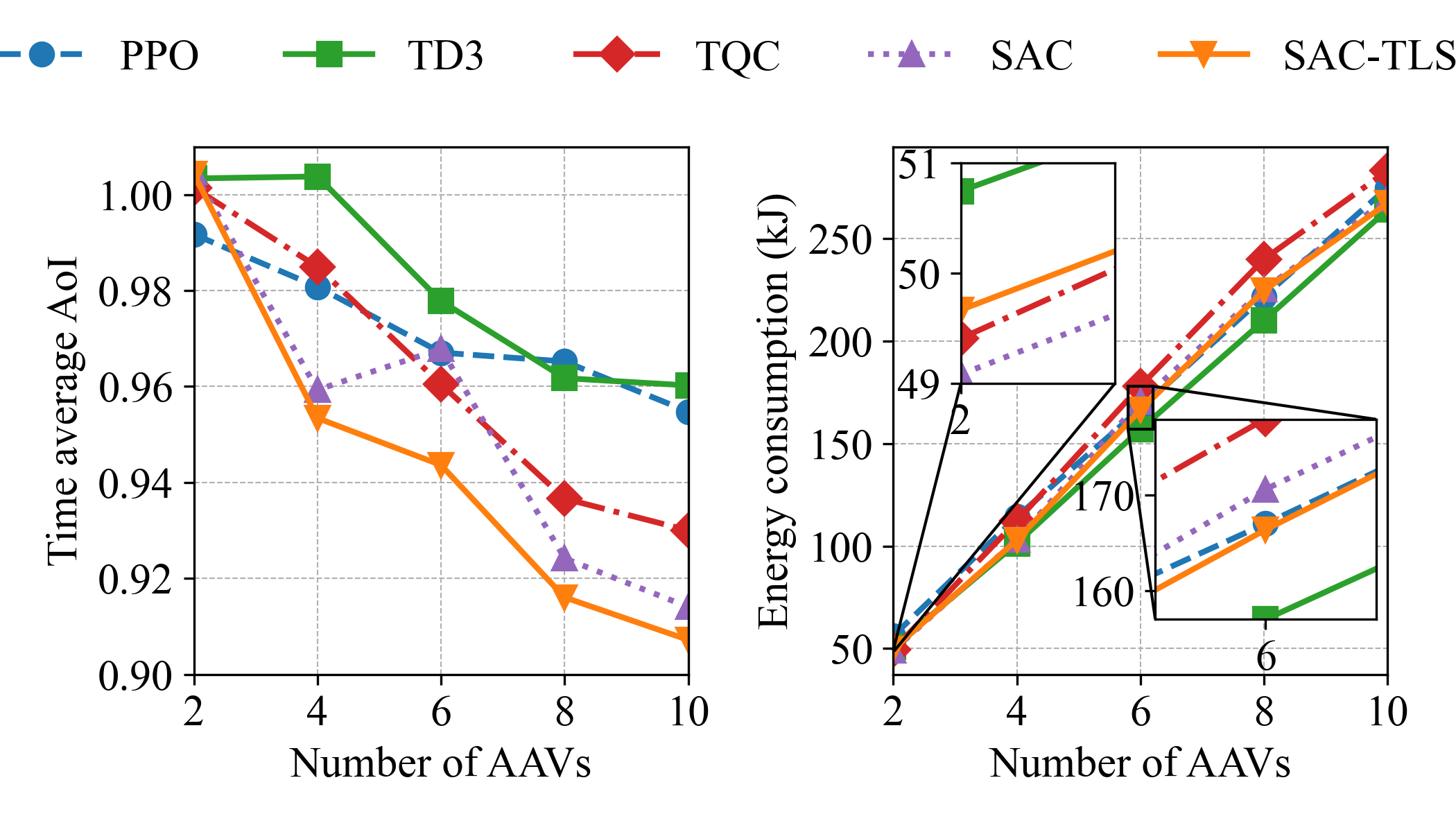}
    \caption{The impacts for different AAV numbers.}
    \label{fig:AAV_num}
\end{figure}
\subsection{Optimization Performance of SAC-TLS and Benchmark Algorithms}
\par Fig.~\ref{fig:Simulation results}(a) illustrates the cumulative rewards for each episode of SAC-TLS in comparison to other benchmark algorithms. As can be seen, the SAC-TLS algorithm outperforms other algorithms with higher rewards, faster convergence, and better stability. The reason may be attributed to following factors. First, our temporal sequence input enables the algorithm to capture evolving environment dynamics. This allows the AAVs to anticipate changes and make proactive decisions. Second, LNGRU stabilizes learning by effectively handling long-term dependencies and preventing gradient issues. This ensures efficient policy updates over extended time horizons. Finally, the SE block focuses on SNs with highly AoI, and optimizes AAV actions to reduce unnecessary exploration. In this case, these components enable SAC-TLS to learn more efficiently, which leads to faster convergence by streamlining decision-making and optimizing both data collection and energy consumption. 

\par Fig.~\ref{fig:Simulation results}(b) provides the comparative results of the time average AoI of SNs and energy consumption of AAVs among five DRL methods and greedy method. We can observe that the SAC-TLS algorithm achieves the lowest AoI while maintaining a competitive energy consumption. This indicates the effectiveness of SAC-TLS in balancing timely data collection with efficient resource utilization.  In contrast, the greedy algorithm shows higher AoI and also exhibits higher energy consumption compared to the DRL methods, which reflects its myopic decision-making strategy that prioritizes immediate rewards at the cost of long-term efficiency. 
While TD3 achieves the lowest energy consumption, its AoI performance is suboptimal, likely due to its more conservative action selection approach. Overall, these results emphasize that while greedy algorithms provide simple, immediate solutions, DRL methods, particularly the proposed SAC-TLS, successfully establish an optimal trade-off between the energy consumption and AoI, thereby highlighting its robustness in dynamic environments. 

\begin{figure}
    \centering
    \includegraphics[width=0.9\linewidth]{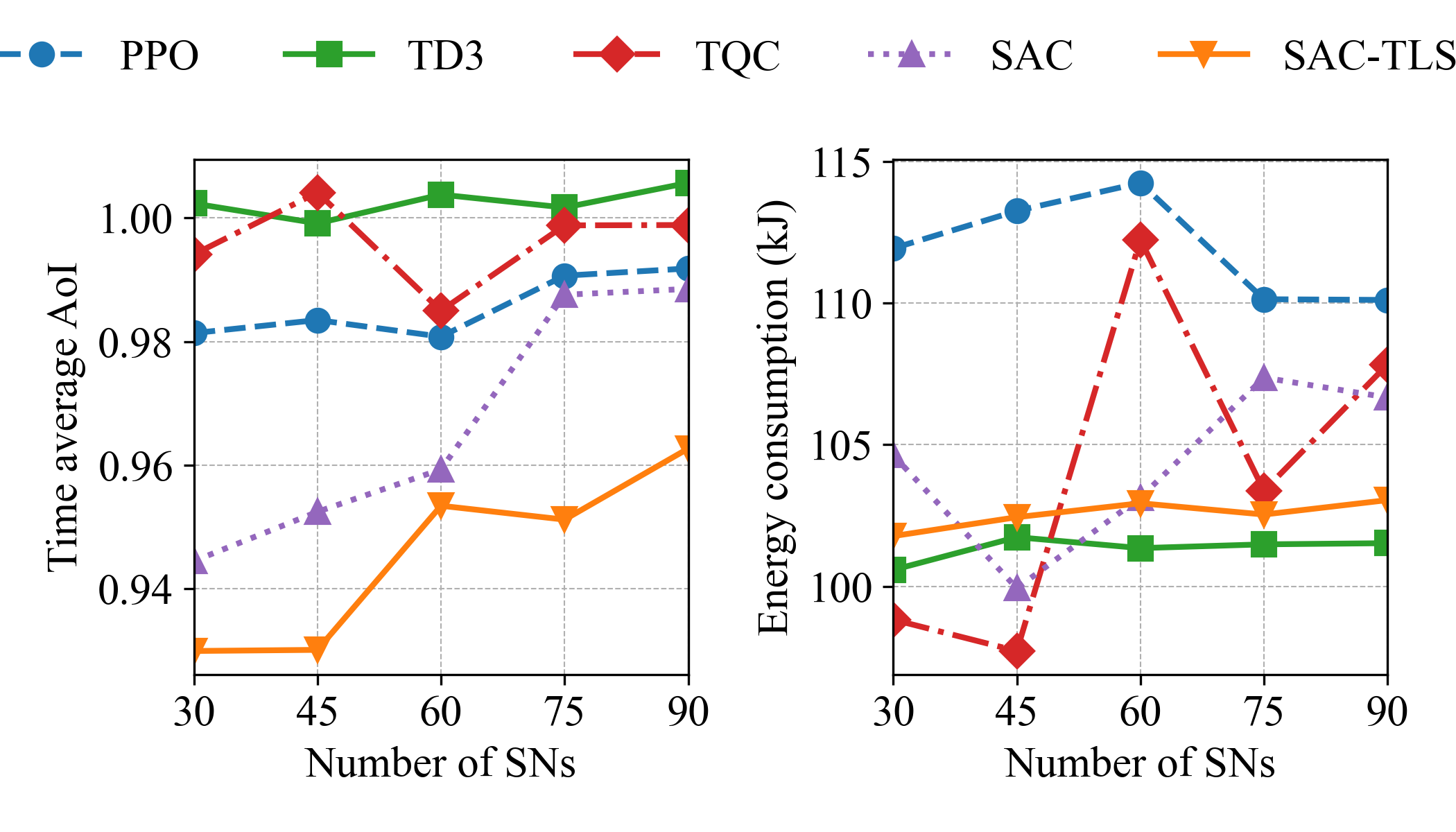}
    \caption{The impacts for different SN numbers.}
    \label{fig:sensor_num}
\end{figure}
\subsection{Impacts of Simulation Parameters}
\par Fig.~\ref{fig:AAV_num} shows the impact of different numbers of AAVs on our SAC-TLS and other benchmark algorithms. Specifically, we observe that as the number of AAVs increases, the average AoI decreases for all algorithms, although the rate of improvement varies. This trend can be attributed to the increased availability of AAVs, which improves the overall data collection and information delivery process and reduces the time delay in updating information. Among the algorithms, SAC-TLS consistently achieves the lowest AoI. On the other hand, we observe that energy consumption increases with the number of AAVs for all algorithms. This is expected, as more AAVs contribute to higher energy usage due to the additional communication and processing tasks. Overall, our SAC-TLS successfully establishes an optimal trade-off between the AoI and energy consumption as the number of AAVs varies. 

\par Fig.~\ref{fig:sensor_num} illustrates the impact of different numbers of SNs on SAC-TLS and other benchmark algorithms. We can observe that the time average AoI generally decreases as the number of SNs increases. This is because more SNs allow the system to gather and transmit more frequent updates, which reduces the staleness of information. The proposed SAC-TLS consistently achieves the lowest AoI, which reflects its superior ability to manage the trade-off between the exploration and exploitation. Moreover, the energy consumption remains within a certain range for the same number of AAVs, which indicates the AAVs have learned to fly around the speed at which the energy consumption is the lowest. 
\par In summary, the proposed SAC-TLS algorithm consistently outperforms other algorithms in terms of total rewards, time average AoI, and energy consumption in the proposed scenarios. 

\begin{figure}
    \centering
    \includegraphics[width=0.97\linewidth]{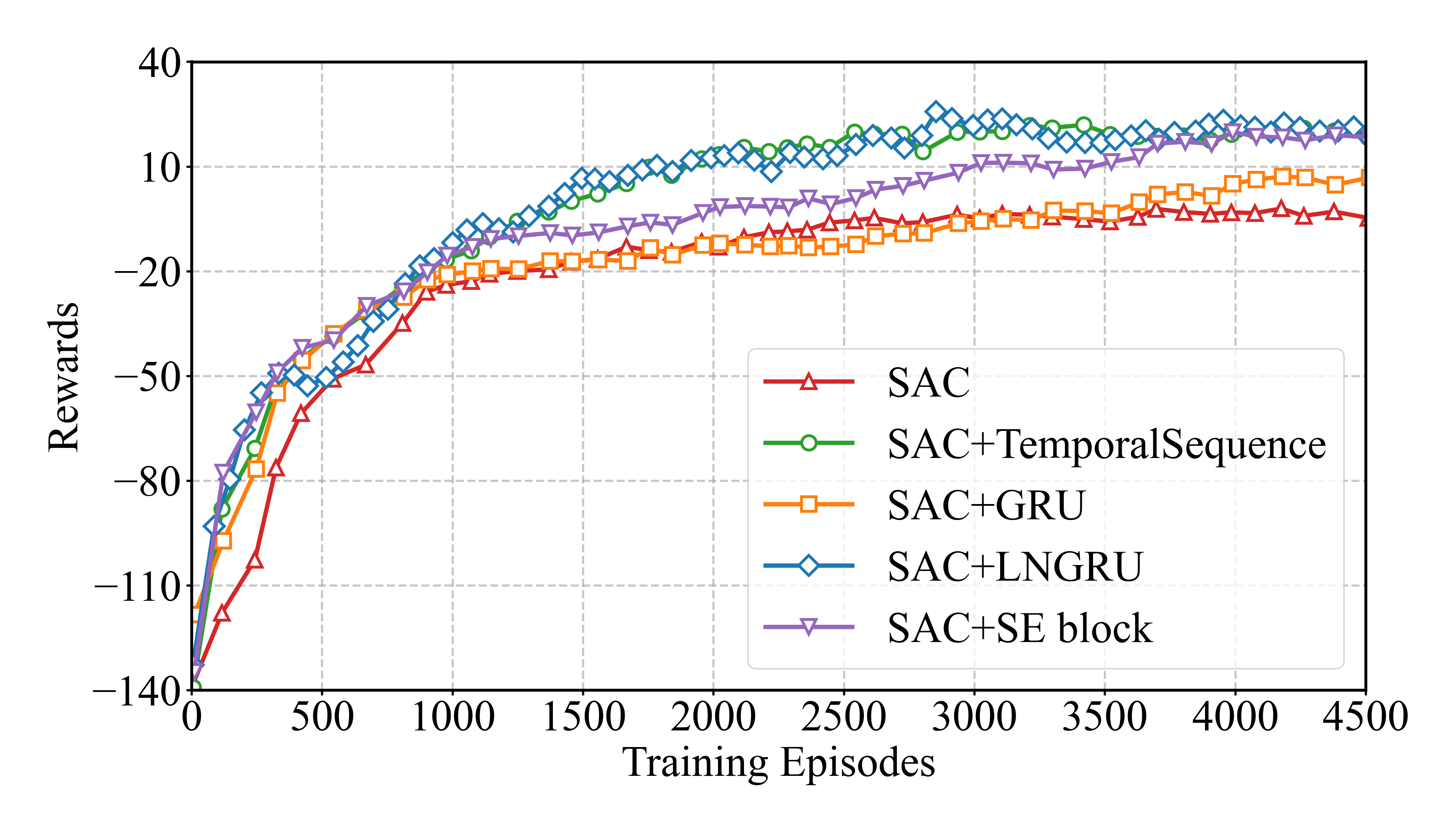}
    \caption{Ablation simulation results.}
    \label{fig:xiaorong}
\end{figure}
\subsection{Ablation Simulation Results} 
\par This section presents comprehensive simulation results that demonstrate the performance enhancements achieved through each individual improvement to the original algorithm. As illustrated in Fig.~\ref{fig:xiaorong}, we compare the standard SAC algorithm against four enhanced variants to validate the effectiveness of our proposed improvements. In addition, the comparison between the SAC with GRU and SAC with LNGRU is specifically designed to verify whether the layer normalization modification to the standard GRU architecture provides meaningful benefits in reinforcement learning contexts. The figure clearly shows that all enhanced variants outperform standard SAC, particularly the LNGRU-enhanced and temporal sequence-enhanced implementations, which achieve the highest final rewards and fastest convergence rates. 

\par The performance differences among these algorithmic variants can be attributed to their distinct enhancement mechanisms. The temporal sequence integration enables the model to effectively capture time dependencies by incorporating sequential information, which leads the model to a better understanding of the relationships between long-term states and actions. 
In the recurrent architecture comparison, while SAC with standard GRU provides modest improvements over baseline SAC, the incorporation of LNGRU delivers substantially greater benefits in both convergence speed and final reward levels.  The reason may be that GRU networks often suffer from training instability, gradient explosion, and vanishing gradient problems when dealing with complex temporal dependencies. LNGRU directly addresses these limitations by incorporating layer normalization, which stabilizes the hidden state distributions during training and effectively mitigates gradient issues. Meanwhile, SE block optimization focuses on important features through a channel attention mechanism that selectively emphasizes critical input features, allowing the model to adaptively recalibrate channel-wise feature responses and concentrate computational resources on the most informative aspects of the input data.

\par Overall, the ablation simulation confirms that each proposed enhancement contributes significantly to the algorithm performance. These results reveal that time sequence improves the temporal dependency of the SAC, LNGRU stabilizes the gradients of the algorithm, and SE block enhances feature extraction in the SAC. 

\begin{figure}
    \centering
    \includegraphics[width=0.9\linewidth]{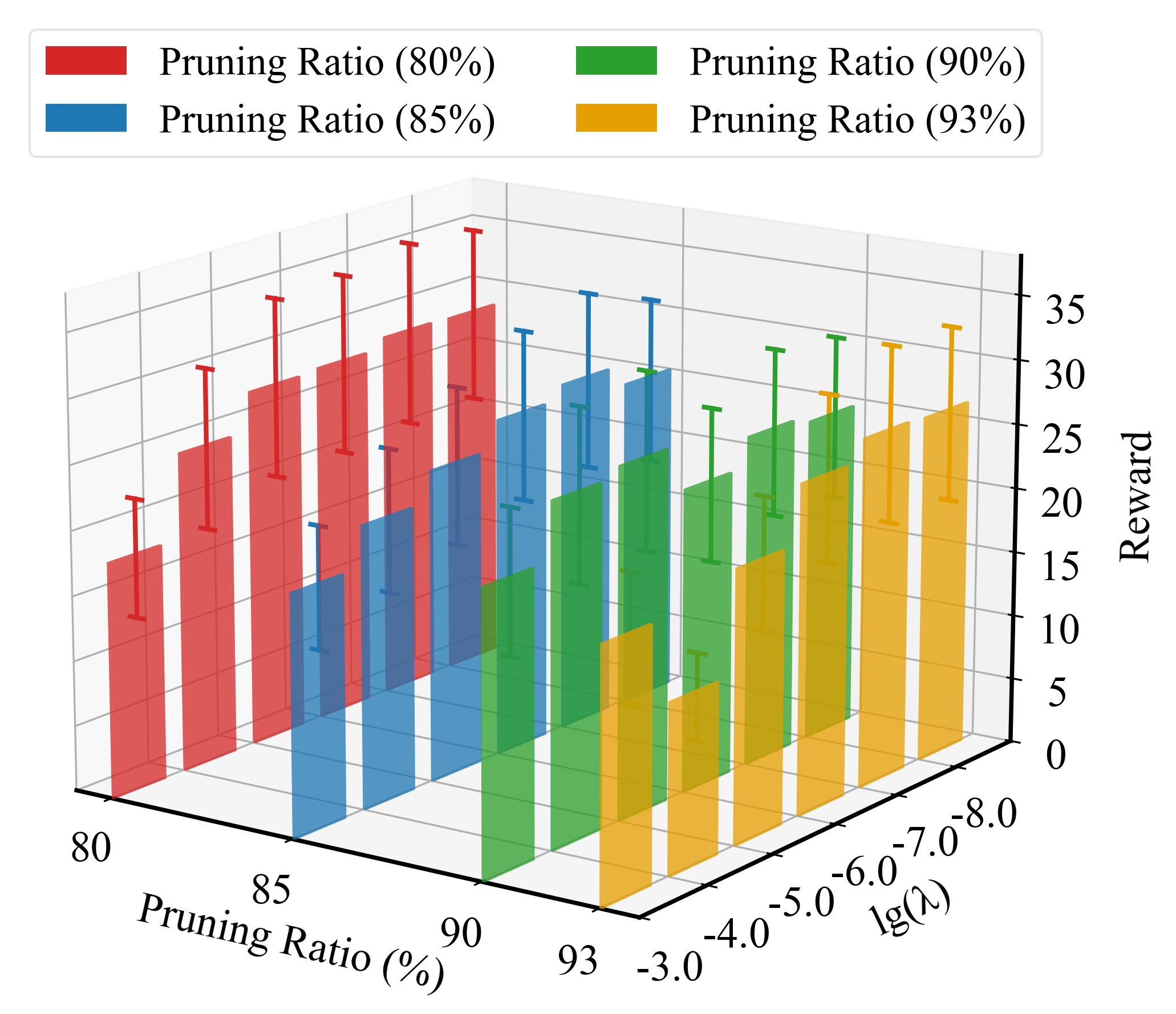}
        \caption{Performance Tradeoff under Structured Pruning and Regularization in SAC-TLS.}
    \label{fig:model reduction}
\end{figure}
\subsection{Compression Performance Tradeoffs in SAC-TLS}
Fig.~\ref{fig:model reduction} illustrates the performance of our SAC-TLS model after applying dynamic structured pruning (DSP) method~\cite{Su2024}. The 3D visualization captures the interaction between pruning ratios ranging from 80\% to 93\%, the regularization strength represented by $\lg \lambda$, and the corresponding reward values. The results reveal a clear trade-off between model compression and performance. As the pruning ratio increases, the reward tends to decrease, indicating a loss in representational capacity due to parameter reduction. However, this degradation is strongly influenced by the choice of the regularization parameter $\lambda$. When $\lg \lambda$ lies between $-7.0$ and $-6.0$, the model is able to retain relatively high performance even under aggressive pruning. This outcome suggests that well-tuned regularization plays a critical role in preserving essential network structures during compression. In particularly, under an optimal regularization setting around $\lg \lambda = -6.0$, the model achieves compression ratios as high as 90\%, with only a marginal drop of approximately 20\% in performance, which is beneficial for efficiently deploying SAC-TLS in resource-constrained environments while maintaining good performance.

\section{Conclusion}\label{sec:conclusion}
\par We have investigated an AAV-assisted AoI-sensitive data forwarding system, where distributed beamforming has been used to overcome long-range transmission barriers between SNs and the BS. To minimize AoI and energy consumption, we have formulated a joint optimization problem that integrates AAV trajectory planning with communication scheduling. The problem has proved highly challenging due to its dynamic and non-convex nature with time-coupled constraints and complex cross-tier dependencies between multi-tiers. To address this, we have proposed SAC-TLS, an enhanced SAC algorithm with temporal sequence input, LNGRU, and SE blocks. These modules have improved the ability of model to learn temporal patterns, maintain stability, and focus on critical features. Simulation results have demonstrated that the proposed SAC-TLS achieves faster convergence and superior performance compared to various baseline algorithms. Moreover, parameter impact analysis has confirmed its robustness under varying settings. Ablation studies have demonstrated the critical role of LNGRU and SE block. Furthermore, model compression experiments have shown that SAC-TLS maintains strong performance even under high pruning rates. Future work will extend this framework to broader AAV-based scenarios and explore generative AI techniques to further improve decision-making under uncertainty.

\ifCLASSOPTIONcaptionsoff
  \newpage
\fi

\bibliographystyle{IEEEtran}  
\bibliography{aoi}     

\begin{thebibliography}{10}
\providecommand{\url}[1]{#1}
\csname url@samestyle\endcsname
\providecommand{\newblock}{\relax}
\providecommand{\bibinfo}[2]{#2}
\providecommand{\BIBentrySTDinterwordspacing}{\spaceskip=0pt\relax}
\providecommand{\BIBentryALTinterwordstretchfactor}{4}
\providecommand{\BIBentryALTinterwordspacing}{\spaceskip=\fontdimen2\font plus
\BIBentryALTinterwordstretchfactor\fontdimen3\font minus \fontdimen4\font\relax}
\providecommand{\BIBforeignlanguage}[2]{{%
\expandafter\ifx\csname l@#1\endcsname\relax
\typeout{** WARNING: IEEEtran.bst: No hyphenation pattern has been}%
\typeout{** loaded for the language `#1'. Using the pattern for}%
\typeout{** the default language instead.}%
\else
\language=\csname l@#1\endcsname
\fi
#2}}
\providecommand{\BIBdecl}{\relax}
\BIBdecl

\bibitem{Lang2025}
Z.~Lang, G.~Liu, G.~Sun, J.~Li, Z.~Sun, J.~Wang, and V.~C.~M. Leung, ``{A}o{I}-sensitive data forwarding with distributed beamforming in {UAV}-assisted {I}o{T},'' \emph{Co{RR}}, vol. abs/2502.09038, 2025.

\bibitem{Huang2024}
C.~Huang, S.~Fang, H.~Wu, Y.~Wang, and Y.~Yang, ``Low-altitude intelligent transportation: System architecture, infrastructure, and key technologies,'' \emph{J. Ind. Inf. Integr.}, vol.~42, p. 100694, 2024.

\bibitem{feng2025networked}
Y.~Feng, C.~Zhao, H.~Luo, F.~Gao, F.~Liu, and S.~Jin, ``Networked {ISAC} based {UAV} tracking and handover towards low-{A}ltitude economy,'' \emph{IEEE Transactions on Wireless Communications}, 2025.

\bibitem{He2024}
L.~He, G.~Sun, Z.~Sun, P.~Wang, J.~Li, S.~Liang, and D.~Niyato, ``An online joint optimization approach for qoe maximization in uav-enabled mobile edge computing,'' in \emph{Proc. {IEEE} {INFOCOM} 2024}, 2024, pp. 101--110.

\bibitem{Xiao2024}
Y.~Xiao, Z.~Ye, M.~Wu, H.~Li, M.~Xiao, M.~Alouini, A.~Al{-}Hourani, and S.~Cioni, ``Space-{A}ir-{G}round integrated wireless networks for 6g: Basics, key technologies, and future trends,'' \emph{{IEEE} J. Sel. Areas Commun.}, vol.~42, no.~12, pp. 3327--3354, 2024.

\bibitem{Sun2024f}
Z.~Sun, G.~Sun, L.~He, F.~Mei, S.~Liang, and Y.~Liu, ``A two time-{S}cale joint optimization approach for {UAV}-assisted {MEC},'' in \emph{{IEEE} {INFOCOM}}.\hskip 1em plus 0.5em minus 0.4em\relax {IEEE}, 2024, pp. 91--100.

\bibitem{Xie2025}
W.~Xie, G.~Sun, B.~Liu, J.~Li, J.~Wang, H.~Du, D.~Niyato, and D.~I. Kim, ``Joint optimization of uav-carried {IRS} for urban low altitude mm{W}ave communications with deep reinforcement learning,'' \emph{CoRR}, vol. abs/2501.02787, 2025.

\bibitem{Sun2024c}
G.~Sun, Y.~Wang, Z.~Sun, Q.~Wu, J.~Kang, D.~Niyato, and V.~C.~M. Leung, ``Multi-objective optimization for multi-uav-assisted mobile edge computing,'' \emph{{IEEE} Trans. Mob. Comput.}, vol.~23, no.~12, pp. 14\,803--14\,820, 2024.

\bibitem{Sun2024d}
G.~Sun, X.~Zheng, Z.~Sun, Q.~Wu, J.~Li, Y.~Liu, and V.~C.~M. Leung, ``{UAV}-{E}nabled secure communications via collaborative beamforming with imperfect eavesdropper information,'' \emph{{IEEE} Trans. Mob. Comput.}, vol.~23, no.~4, pp. 3291--3308, 2024.

\bibitem{Sun2024e}
Z.~Sun, G.~Sun, Y.~Liu, J.~Wang, and D.~Cao, ``{BARGAIN-MATCH:} {A} game theoretical approach for resource allocation and task offloading in vehicular edge computing networks,'' \emph{{IEEE} Trans. Mob. Comput.}, vol.~23, no.~2, pp. 1655--1673, 2024.

\bibitem{Bai2025}
Z.~Bai, J.~Shi, Z.~Li, M.~Li, and K.~Chen, ``Rule-guided {DRL} for {UAV}-{A}ssisted wireless sensor networks with no-{F}ly zones safety,'' \emph{{IEEE} Trans. Cogn. Commun. Netw.}, vol.~11, no.~2, pp. 1268--1280, 2025.

\bibitem{Sun2024b}
G.~Sun, L.~He, Z.~Sun, Q.~Wu, S.~Liang, J.~Li, D.~Niyato, and V.~C.~M. Leung, ``Joint task offloading and resource allocation in aerial-terrestrial {UAV} networks with edge and fog computing for post-disaster rescue,'' \emph{{IEEE} Trans. Mob. Comput.}, vol.~23, no.~9, pp. 8582--8600, 2024.

\bibitem{Li2024}
J.~Li, G.~Sun, L.~Duan, and Q.~Wu, ``Multi-objective optimization for {UAV} swarm-assisted {I}o{T} with virtual antenna arrays,'' \emph{{IEEE} Trans. Mob. Comput.}, vol.~23, no.~5, pp. 4890--4907, 2024.

\bibitem{Li2024c}
J.~Li, G.~Sun, Q.~Wu, S.~Liang, P.~Wang, and D.~Niyato, ``Two-way aerial secure communications via distributed collaborative beamforming under eavesdropper collusion,'' in \emph{Proc. {IEEE} {INFOCOM} 2024}.\hskip 1em plus 0.5em minus 0.4em\relax {IEEE}, 2024, pp. 331--340.

\bibitem{Zhang2024}
C.~Zhang, G.~Sun, Q.~Wu, J.~Li, S.~Liang, D.~Niyato, and V.~C.~M. Leung, ``{UAV} swarm-enabled collaborative secure relay communications with time-domain colluding eavesdropper,'' \emph{{IEEE} Trans. Mob. Comput.}, vol.~23, no.~9, pp. 8601--8619, 2024.

\bibitem{Li2023}
J.~Li, G.~Sun, H.~Kang, A.~Wang, S.~Liang, Y.~Liu, and Y.~Zhang, ``Multi-{O}bjective optimization approaches for physical layer secure communications based on collaborative beamforming in {UAV} networks,'' \emph{{IEEE/ACM} Trans. Netw.}, vol.~31, no.~4, pp. 1902--1917, 2023.

\bibitem{Liu2024}
S.~Liu, G.~Sun, J.~Li, S.~Liang, Q.~Wu, P.~Wang, and D.~Niyato, ``Uav-enabled collaborative beamforming via multi-agent deep reinforcement learning,'' \emph{{IEEE} Trans. Mob. Comput.}, vol.~23, no.~12, pp. 13\,015--13\,032, 2024.

\bibitem{qin2022aoi}
Z.~Qin, Z.~Wei, Y.~Qu, F.~Zhou, H.~Wang, D.~W.~K. Ng, and C.-B. Chae, ``{AoI}-aware scheduling for air-ground collaborative mobile edge computing,'' \emph{{IEEE} Transactions on Wireless Communications}, vol.~22, no.~5, pp. 2989--3005, 2022.

\bibitem{Zhu2023}
B.~Zhu, E.~Bedeer, H.~H. Nguyen, R.~Barton, and Z.~Gao, ``{UAV} trajectory planning for aoi-minimal data collection in {UAV}-aided {I}o{T} networks by transformer,'' \emph{{IEEE} Trans. Wirel. Commun.}, vol.~22, no.~2, pp. 1343--1358, 2023.

\bibitem{huang2025learning}
Z.~Huang, H.~Chen, B.~Gu, S.~Gong, Z.~Su, and M.~Guizani, ``A learning-based iterative algorithm for {A}o{I}-optimal trajectory planning in {UAV}-assisted {I}o{T} networks,'' \emph{{IEEE} Transactions on Wireless Communications}, 2025.

\bibitem{yang2025oh}
B.~Yang, Y.~Yu, X.~Hao, P.~L. Yeoh, J.~Zhang, L.~Guo, and Y.~Li, ``{OH}-{DRL}: An {A}o{I}-guaranteed energy-efficient approach for {UAV}-assisted {I}o{T} data collection,'' \emph{{IEEE} Transactions on Wireless Communications}, 2025.

\bibitem{zhou2025}
J.~Zhou, M.~Wang, D.~Tian, K.~Qu, G.~Qu, X.~Duan, and X.~Shen, ``Reliability-optimal {UAV}-{A}ssisted mobile edge computing: Joint resource allocation, data transmission scheduling and motion control,'' \emph{{IEEE} Trans. Mob. Comput.}, vol.~24, no.~5, pp. 4217--4234, 2025.

\bibitem{han2022age}
Z.~Han, Y.~Yang, W.~Wang, L.~Zhou, T.~N. Nguyen, and C.~Su, ``Age efficient optimization in uav-aided {VEC} network: A game theory viewpoint,'' \emph{{IEEE} Transactions on Intelligent Transportation Systems}, vol.~23, no.~12, pp. 25\,287--25\,296, 2022.

\bibitem{Yang2023}
Y.~Yang, W.~Wang, L.~Liu, K.~Dev, and N.~M.~F. Qureshi, ``{A}o{I} optimization in the {UAV}-{A}ided traffic monitoring network under attack: {A} stackelberg game viewpoint,'' \emph{{IEEE} Trans. Intell. Transp. Syst.}, vol.~24, no.~1, pp. 932--941, 2023.

\bibitem{Feng2024}
H.~Feng, J.~Wang, Z.~Fang, J.~Chen, and D.~Do, ``Evaluating {A}o{I}-centric {HARQ} protocols for {UAV} networks,'' \emph{{IEEE} Trans. Commun.}, vol.~72, no.~1, pp. 288--301, 2024.

\bibitem{Zhan2024a}
C.~Zhan, H.~Hu, Z.~Liu, J.~Wang, and R.~Fan, ``Interference-{A}ware online optimization for cellular-connected multiple {UAV} networks with energy constraints,'' \emph{{IEEE} Trans. Mob. Comput.}, vol.~23, no.~12, pp. 13\,804--13\,820, 2024.

\bibitem{long2025lyapunov}
Y.~Long, S.~Gong, S.~Sun, G.~C. Lee, L.~Li, and D.~Niyato, ``Lyapunov-guided deep reinforcement learning for semantic-aware aoi minimization in uav-assisted wireless networks,'' \emph{{IEEE} Transactions on Wireless Communications}, 2025.

\bibitem{Han2022a}
R.~Han, Y.~Wen, L.~Bai, J.~Liu, and J.~Choi, ``Age of information aware {UAV} deployment for intelligent transportation systems,'' \emph{{IEEE} Trans. Intell. Transp. Syst.}, vol.~23, no.~3, pp. 2705--2715, 2022.

\bibitem{Dai2022}
Z.~Dai, C.~H. Liu, Y.~Ye, R.~Han, Y.~Yuan, G.~Wang, and J.~Tang, ``Ao{I}-minimal {UAV} crowdsensing by model-based graph convolutional reinforcement learning,'' in \emph{{IEEE} {INFOCOM} 2022 - {IEEE} Conference on Computer Communications, London, United Kingdom, May 2-5, 2022}.\hskip 1em plus 0.5em minus 0.4em\relax {IEEE}, 2022, pp. 1029--1038.

\bibitem{Zheng2024}
X.~Zheng, G.~Sun, J.~Li, S.~Liang, Q.~Wu, M.~Yin, D.~Niyato, and V.~C.~M. Leung, ``Reliable and energy-efficient communications via collaborative beamforming for {UAV} networks,'' \emph{{IEEE} Trans. Wirel. Commun.}, vol.~23, no.~10, pp. 13\,235--13\,251, 2024.

\bibitem{liu2025joint}
Z.~Liu, X.~Liu, W.~Yang, and X.~Zhang, ``Joint sensing and age of information optimization for energy constrained {UAV} assisted integrated sensing, calculation and communication,'' \emph{IEEE Transactions on Wireless Communications}, 2025.

\bibitem{Liu2025}
X.~Liu, H.~Liu, K.~Zheng, J.~Liu, T.~Taleb, and N.~Shiratori, ``Ao{I}-minimal clustering, transmission and trajectory co-design for uav-assisted {WPCN}s,'' \emph{{IEEE} Trans. Veh. Technol.}, vol.~74, no.~1, pp. 1035--1051, 2025.

\bibitem{Samir2022}
M.~Samir, C.~Assi, S.~Sharafeddine, and A.~Ghrayeb, ``Online altitude control and scheduling policy for minimizing aoi in {UAV}-assisted {I}o{T} wireless networks,'' \emph{{IEEE} Trans. Mob. Comput.}, vol.~21, no.~7, pp. 2493--2505, 2022.

\bibitem{Zakeri2024}
A.~Zakeri, M.~Moltafet, M.~Leinonen, and M.~Codreanu, ``Minimizing the {A}o{I} in resource-constrained multi-source relaying systems: Dynamic and learning-based scheduling,'' \emph{{IEEE} Trans. Wirel. Commun.}, vol.~23, no.~1, pp. 450--466, 2024.

\bibitem{Qu2024}
L.~Qu, A.~Huang, and M.~J. Khabbaz, ``Optimization of information freshness in multi-{RIS} cooperative assisted wireless sensor network,'' \emph{{IEEE} Trans. Wirel. Commun.}, vol.~23, no.~11, pp. 16\,332--16\,345, 2024.

\bibitem{Zhao2025}
M.~Zhao, Y.~Xiao, J.~Yao, T.~Wang, J.~Lee, and T.~Q.~S. Quek, ``Up-downlink aoi-driven multi-source data collection in uav-assisted wireless sensor networks,'' \emph{{IEEE} Trans. Wirel. Commun.}, vol.~24, no.~2, pp. 1178--1192, 2025.

\bibitem{chen2025average}
Q.~Chen, S.~Guo, W.~Xu, J.~Li, T.~Shi, H.~Gao, and Z.~Cai, ``Average {A}o{I} minimization with directional charging for wireless-{P}owered network edge,'' \emph{IEEE Transactions on Mobile Computing}, 2025.

\bibitem{Gao2023}
X.~Gao, X.~Zhu, and L.~Zhai, ``{A}o{I}-sensitive data collection in multi-{UAV}-assisted wireless sensor networks,'' \emph{{IEEE} Trans. Wirel. Commun.}, vol.~22, no.~8, pp. 5185--5197, 2023.

\bibitem{Qi2023}
N.~Qi, Z.~Huang, W.~Sun, S.~Jin, and X.~Su, ``Coalitional formation-based group-buying for {UAV}-enabled data collection: An auction game approach,'' \emph{{IEEE} Trans. Mob. Comput.}, vol.~22, no.~12, pp. 7420--7437, 2023.

\bibitem{Hoang2024}
T.~M. Hoang, B.~C. Nguyen, T.~T.~H. Le, X.~N. Tran, and P.~T. Hiep, ``Finite block length {NOMA} {MU} pairing {UAV}-enable system: Performance analysis and optimization,'' \emph{{IEEE} Trans. Mob. Comput.}, vol.~23, no.~10, pp. 9804--9820, 2024.

\bibitem{Zhan2024}
C.~Zhan, H.~Hu, J.~Wang, Z.~Liu, and S.~Mao, ``Tradeoff between age of information and operation time for {UAV} sensing over multi-cell cellular networks,'' \emph{{IEEE} Trans. Mob. Comput.}, vol.~23, no.~4, pp. 2976--2991, 2024.

\bibitem{Reddy2025}
Y.~A.~K. Reddy and T.~G. Venkatesh, ``Proactive obsolete packet management based analysis of age of information for {LCFS} heterogeneous queueing system,'' \emph{{IEEE} Trans. Mob. Comput.}, vol.~24, no.~3, pp. 1513--1529, 2025.

\bibitem{Zheng2024a}
H.~Zheng, K.~Xiong, M.~Sun, H.~Wu, Z.~Zhong, and X.~Shen, ``Maximizing age-energy efficiency in wireless powered industrial {I}o{E} networks: {A} dual-layer {DQN}-based approach,'' \emph{{IEEE} Trans. Wirel. Commun.}, vol.~23, no.~2, pp. 1276--1292, 2024.

\bibitem{Li2024b}
J.~Li, X.~Wang, J.~Wu, and Z.~Ning, ``Intelligent scheduling of uavs and sensors for information age minimization at wireless powered internet of things,'' in \emph{Proc. {IEEE} {CSCWD}}, W.~Shen, J.~A. Barth{\`{e}}s, J.~Luo, T.~Qiu, X.~Zhou, J.~Zhang, H.~Zhu, K.~Peng, T.~Xu, and N.~Chen, Eds., 2024, pp. 3243--3248.

\bibitem{zhang2025aoi}
G.~Zhang, X.~Wei, X.~Tan, Z.~Han, and G.~Zhang, ``{A}o{I} minimization based on deep reinforcement learning and matching game for {I}o{T} information collection in {SAGIN},'' \emph{IEEE Transactions on Communications}, 2025.

\bibitem{Huang2025}
Z.~Huang, W.~Wu, K.~Wu, H.~Yuan, C.~Fu, F.~Shan, J.~Wang, and J.~Luo, ``{LI2:} {A} new learning-based approach to timely monitoring of points-of-interest with {UAV},'' \emph{{IEEE} Trans. Mob. Comput.}, vol.~24, no.~1, pp. 45--61, 2025.

\bibitem{Zhang2025}
C.~Zhang, Y.~Zou, Z.~Zhang, D.~Yu, J.~T. G{\'{o}}mez, T.~Lan, F.~Dressler, and X.~Cheng, ``Distributed age-of-information scheduling with {NOMA} via deep reinforcement learning,'' \emph{{IEEE} Trans. Mob. Comput.}, vol.~24, no.~1, pp. 30--44, 2025.

\bibitem{Li2024a}
J.~Li, G.~Sun, Q.~Wu, D.~Niyato, J.~Kang, A.~Jamalipour, and V.~C.~M. Leung, ``Collaborative ground-space communications via evolutionary multi-objective deep reinforcement learning,'' \emph{{IEEE} J. Sel. Areas Commun.}, vol.~42, no.~12, pp. 3395--3411, 2024.

\bibitem{Long2022}
Y.~Long, W.~Zhang, S.~Gong, X.~Luo, and D.~Niyato, ``{A}o{I}-aware scheduling and trajectory optimization for multi-{UAV}-assisted wireless networks,'' in \emph{Proc. {IEEE} {GLOBECOM}}, 2022, pp. 2163--2168.

\bibitem{Wang2023}
F.~Wang, D.~Jiang, Z.~Wang, and S.~Mumtaz, ``Service continuity based data delivery optimization in satellite-terrestrial networks,'' \emph{{IEEE} Trans. Veh. Technol.}, vol.~72, no.~10, pp. 13\,604--13\,617, 2023.

\bibitem{Sun2024a}
A.~Sun, C.~Sun, J.~Du, C.~Chen, C.~Huang, and J.~Sui, ``Aoi optimization for uav-assisted wireless sensor networks,'' in \emph{Proc. {IEEE} {ICC}}, 2024, pp. 1487--1492.

\bibitem{Haarnoja2018}
T.~Haarnoja, A.~Zhou, P.~Abbeel, and S.~Levine, ``Soft actor-critic: Off-policy maximum entropy deep reinforcement learning with a stochastic actor,'' in \emph{Proc. {PMLR} {ICML}}, vol.~80, 2018, pp. 1856--1865.

\bibitem{LeiBa2016}
J.~Lei~Ba, J.~R. Kiros, and G.~E. Hinton, ``Layer normalization,'' \emph{ArXiv e-prints}, pp. arXiv--1607, 2016.

\bibitem{hu2018squeeze}
J.~Hu, L.~Shen, and G.~Sun, ``Squeeze-and-excitation networks,'' in \emph{Proc. {IEEE} {CVPR}}, 2018, pp. 7132--7141.

\bibitem{Fujimoto2018}
S.~Fujimoto, H.~van Hoof, and D.~Meger, ``Addressing function approximation error in actor-critic methods,'' ser. Proceedings of Machine Learning Research, vol.~80.\hskip 1em plus 0.5em minus 0.4em\relax {PMLR}, 2018, pp. 1582--1591.

\bibitem{Schulman2017}
J.~Schulman, F.~Wolski, P.~Dhariwal, A.~Radford, and O.~Klimov, ``Proximal policy optimization algorithms,'' \emph{Co{RR}}, vol. abs/1707.06347, 2017.

\bibitem{Kuznetsov2020}
A.~Kuznetsov, P.~Shvechikov, A.~Grishin, and D.~P. Vetrov, ``Controlling overestimation bias with truncated mixture of continuous distributional quantile critics,'' in \emph{Proc. {PMLR} {ICML}}, vol. 119, 2020, pp. 5556--5566.

\bibitem{Lillicrap2016}
T.~P. Lillicrap, J.~J. Hunt, A.~Pritzel, N.~Heess, T.~Erez, Y.~Tassa, D.~Silver, and D.~Wierstra, ``Continuous control with deep reinforcement learning,'' in \emph{Proc. {ICLR}}, 2016.

\bibitem{Su2024}
W.~Su, Z.~Li, M.~Xu, J.~Kang, D.~Niyato, and S.~Xie, ``Compressing deep reinforcement learning networks with a dynamic structured pruning method for autonomous driving,'' \emph{{IEEE} {T}rans. {V}eh. {T}echnol.}, vol.~73, no.~12, pp. 18\,017--18\,030, 2024.

\end{thebibliography}

\end{document}